\documentclass{desyproc}

\begin{document}
%------------------------------------
\title{Beauty and charm results from $B$-factories}

%for single authors the superscripts are optional
\author{{\slshape Bo\v stjan Golob}\\[1ex]
Faculty of Mathematics and Physics, University of Ljubljana,
  Jadranska 19, 1000 Ljubljana, Slovenia\\
Jo\v zef Stefan Institute, Jamova 39, 1000 Ljubljana, SLovenia}

% if the proceedings are available online (e.g. at Indico)
% please enter the contribution ID or file_name below for the DOI
%\contribID{32}
%\contribID{golob\_bostjan}

% TO THE CONFERENCE EDITORS: 
% please update the following information      
% before sending the template to the authors
% \confID{800}  % if the conference is on Indico uncomment this line
\desyproc{DESY-PROC-2008-xx}
\acronym{HQP08} % if you want the Acronym in the page footer uncomment this line
\doi  % if there is an online version we will register DOIs

\maketitle

\begin{abstract}
We present the proceedings of the lectures given at the 2008 Helmholtz International Summer
School Heavy Quark Physics at the Bogoliubov Laboratory of Theoretical
Physics in Dubna. 
In two lectures we present recent results from the existing
$B$-factories experiments, Belle and BaBar. The discussed topics 
include short phenomenological motivation, experimental methods and
results on $B$ meson oscillations, selected rare $B$ meson decays
(leptonic, $b\to s\gamma$ and $b\to s\ell^+\ell^-$), mixing and
$CP$ violation in the system of $D^0$ mesons, and leptonic decays of $D_s$
mesons. 
\end{abstract}

\section{Introduction}
\label{sec_1}

The lectures presented in this paper are a part of the $B$-factories
lectures prepared in collaboration with A.J. Bevan (also given in the
proceedings of the school, \cite{bevan_proc}). To obtain an approximate overview of
recent results on flavour physics arising from Belle and BaBar both
sets of presentations (each composed of two one-hour lectures) should
be consulted. 

The lectures presented here include - beside the experimental methods
and results - some short phenomenological sketches of motivation
and/or interpretation of individual measurements. The author, being an
experimentalist, should warn the reader that some examples of
phenomenological interpretation are simplified and that serious
theoretical treatment requires consultation of references given in the
text. Examples are thus to be treated with a grain of salt; to quote the
famous poet: {\it "It is a curious fact that
  people are never so trivial as when they take themselves seriously." 
(O. Wilde, 1854 - 1900)}.  

A large majority of results presented in the lectures arise from the
measurements performed with the two experiments taking data at the
$B$-factories, $e^+e^-$ asymmetric colliders running at the
center-of-mass (CM) energy
$\sqrt{s}=m_{\Upsilon(4S)}c^2$~\footnote{Here and in the following we adopt a
  notation where $m_X$ represents a nominal mass value of particle
  $X$. If we refer to the reconstructed invariant mass of a system
  $Y$ we use the notation $m(Y)$.}. $\Upsilon(4S)$,
a $b\bar{b}$ bound state with a mass just above the threshold for
$\Upsilon(4S)\to B\bar{B}$ decay, is a copious source of $B$
meson pairs. Mesons are produced almost at rest in the CM system, but
since the electron beam has an energy higher than the positron one,
they are boosted and decay time dependent measurements of meson decays
are thus possible. The Belle
detector \cite{belle_det}, operating at the KEKB collider \cite{kekb} in Tsukuba, Japan, has so far
recorded an integrated luminosity of around 860~fb$^{-1}$,
roughly corresponding to $950\times 10^6$ pairs of $B$ mesons
\footnote{Both, $B^0\bar{B}^0$ and $B^+B^-$ pairs are produced, at 
  approximately the same rate.}. The BaBar detector \cite{babar_det}
  at the PEP-II collider in Stanford, USA, has recorded
  around 550~fb$^{-1}$ of data. 

\begin{figure}[hb]
\centerline{\includegraphics[width=0.7\textwidth]{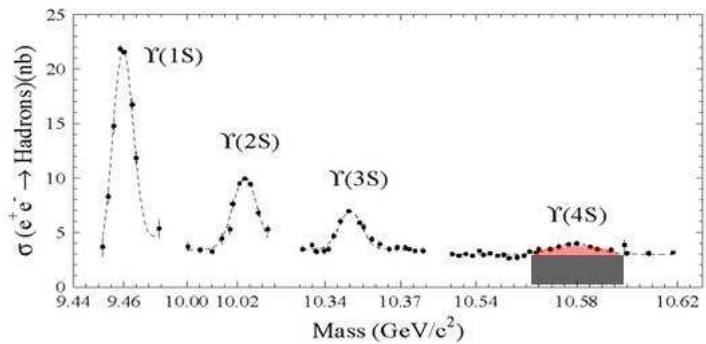}}
\caption{Cross-section for production of hadrons in $e^+e^-$
  collisions as a function of $\sqrt{s}$. The resonant production of
  $\Upsilon(4S)\to B\bar{B}$ is represented as the light shaded area, and
  the continuum $e^+e^-\to\gamma^\ast\to q\bar{q}$ as the dark shaded
  area. }
\label{fig_1}
\end{figure}

Beside the production of $B$ meson pairs from the $\Upsilon(4S)$ other
processes take place in $e^+e^-$ collisions at the given CM energy. For
the subject of the lectures the most important is the continuum
production of $c\bar{c}$ quark pairs, arising in $e^+e^-\to
\gamma^\ast\to c\bar{c}$. This is sketched in Fig. \ref{fig_1}, where the
cross-section for hadron production in electron-positron collisions
is plotted as a function of the CM collision energy. 
The cross-section for the production of $c\bar{c}$ pairs is larger
than the one for the $B$ meson production, at the integrated luminosity of
KEKB it corresponds to around $1.1\times 10^9$ produced pairs of
charmed hadrons. 

In the course of the lectures we will mention also some related results from
experiments other than $B$-factories, specifically the ones from the
CDF-II experiment at Tevatron \cite{cdf_det}, recording data in
$p\bar{p}$ collisions, and Cleo-c experiment \cite{cleoc_det} at 
the $e^+e^-$ collider CESR, running at the $D\bar{D}$ meson pair production threshold. All these
experiments provide for a truly diverse experimental environment to
study various aspects of heavy flavour physics. {\it "We all live with
the objective of being happy; our lives are all different and yet the
same." (A. Frank, 1929 - 1945)}. 

\section{Lecture I}
{\it "Never loose an opportunity of seeing anything beautiful, for
  {\bf beauty} is God's handwriting." (R.W. Emerson, 1803-1882)} 

\subsection{$B$ meson oscillations}
\label{sec_21}

The mixing of neutral mesons, that is the transition of a neutral meson
$P^0$ into its
antiparticle and vice-versa, appears as a consequence of states of
definite flavour ($P^0$, $\bar{P}^0$) being a linear superposition of
the eigenstates of an effective Hamiltonian (states of a simple exponential time evolution) $P_{1,2}$:
\begin{equation}
|P_{1,2}\rangle=p|P^0\rangle\pm q |\bar{P}^0\rangle~~.
\label{eq_1}
\end{equation}
For a thorough derivation of the equations describing the oscillations
of mesons the reader is
advised to follow \cite{pdg_cpv}. 

\begin{figure}[hb]
\centerline{\includegraphics[width=0.8\textwidth]{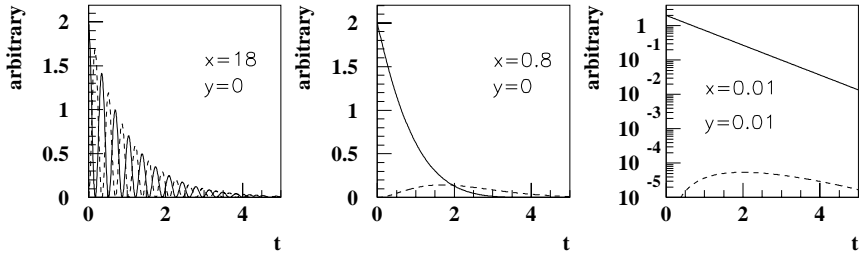}}
\caption{Probability for an initially produced meson $P^0$ to decay at
  time $t$ as $P^0$ (full curve) or $\bar{P}^0$ (dashed 
  curve). Qualitatively the left example roughly corresponds to the case of
  $B_s^0$ mesons, the middle one to the case of $B^0$ mesons and the
  right one to the case of $D^0$ mesons. Note the logarithmic scale on
  the right plot.}
\label{fig_2}
\end{figure}

While the mass eigenstates have a simple time evolution, the time
dependent decay rate of flavour eigenstates depends on the mixing
parameters $x$ and $y$, expressed in terms of the mass and
width difference of $P_{1,2}$ as $x=(m_1-m_2)/\bar{\Gamma}$ and 
$y=(\Gamma_1-\Gamma_2)/2\bar{\Gamma}$. $\bar{\Gamma}$ is the average
decay width of the two mass eigenstates. The decay rate of a state
initially produced as a $P^0$ is 
\begin{eqnarray} 
\nonumber
&&\frac{d\Gamma(P^0\to f)}{dt}=e^{-t}\bigl[\bigl(\vert
A_f\vert^2+\vert\frac{q}{p}\bar{A}_f\vert^2\bigr)\cosh{yt}+ 
 \bigl(\vert
 A_f\vert^2-\vert\frac{q}{p}\bar{A}_f\vert^2\bigr)\cos{xt}\\
&&+2\Re{\bigl(\frac{q}{p}A_f^\ast\bar{A}_f\bigr)}\sinh{yt}-
2\Im{\bigl(\frac{q}{p}A_f^\ast\bar{A}_f\bigr)}\sin{xt}\bigr]~.
\label{eq_2}
\end{eqnarray}
In the above equation $t$ is a dimensionless decay time, defined in
terms of a proper decay time $t^\prime$ as
$t=t^\prime\bar{\Gamma}$. The notation $A_f,~\bar{A}_f$ is used to represent 
instantaneous amplitudes for $P^0\to f$ and $\bar{P}^0\to f$
decays. 
It is obvious from Eq. (\ref{eq_2}) that using the decay time
distribution of experimentally accessible flavour eigenstates one can determine the mixing
parameters $x$ and $y$. Moreover, the effect of the mixing parameters on
$d\Gamma/dt$ depends on the chosen decay channel ($A_f,~\bar{A}_f$). 
The decay time distribution of an initially produced $\bar{P}^0$ is
obtained from Eq. (\ref{eq_2}) by replacing
$A_f\leftrightarrow\bar{A}_f$ and $q/p\to p/q$. 
The decay time
distributions for decays to conjugated final state $\bar{f}$ are 
obtained by a simple $f\to\bar{f}$ transformation.  
The above decay rates are illustrated in Fig. \ref{fig_2} for 
several values of $x$ and $y$. 

The neutral $B$ meson pairs\footnote{We will use notation $B^0$ for
  $B_d^0$ mesons, while for the strange $B$ mesons we will use a
  strict $B_s^0$ notation.} from $\Upsilon(4S)$ decays are produced in a quantum
coherent state with the quantum numbers corresponding to that of the
$\Upsilon(4S)$. Before the coherence is disturbed by a decay of one of
the mesons, the pair is always in a $B^0-\bar{B}^0$ state. The 
decay rates given above are
valid only after the first of the two mesons decays. To be used in
measurements of $B^0$ mesons produced from $\Upsilon(4S)$, the decay time
$t$ in Eq. (\ref{eq_2}) should thus be changed to $\Delta t$, the difference between
the decay times of the first and the second neutral $B$ meson (and the
  exponential factor should include $|\Delta t|$ instead of $t$).  

\begin{figure}[hb]
\centerline{\includegraphics[width=0.7\textwidth]{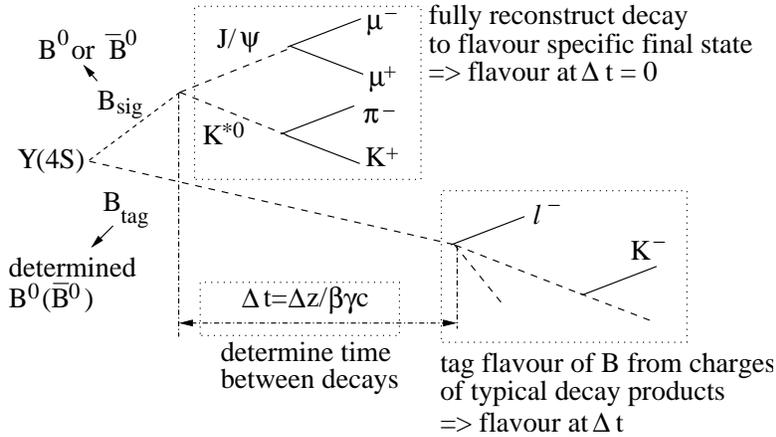}}
\caption{Illustration of the method used to measure the $B^0$
  oscillation frequency $x$.}
\label{fig_3}
\end{figure}

The experimental method of
measuring $B^0$ meson oscillation frequency\footnote{Strictly speaking experiments in $B$ system measure the mass
  difference between the two eigenstates, $\Delta m$. However, since
  the dimensionless mixing parameter $x=\Delta m/\bar{\Gamma}$ can be
  more directly compared for different meson species, we prefer to use
  this. Similarly as for the notation of $B$ mesons, we use $\Delta m$
and $x$ for the $B_d^0$ mesons and $\Delta m_s$ and $x_s$ for $B_s^0$
mesons. In lecture II we will use $x_D$ and $y_D$ to denote
the corresponding mixing parameters in the $D^0$ system.} $x$ relies on a similar method as the one used for measuring the $CP$
violation \cite{bevan_proc}. However, instead of $CP$ specific final states, 
flavour specific final states of $B$ meson decays are used (like
$B^0\to J/\psi K^{\ast 0},~K^{\ast 0}\to K^+\pi^-$), which allow to
determine the flavour of the decaying $B$ meson. 
The method is sketched in
Fig. \ref{fig_3}. The measured $\Delta t$ distribution
deviates from Eq. (\ref{eq_2}) due to several reasons: usage of
flavour specific final state ($\bar{A}_f=A_{\bar{f}}=0$), negligible
decay width difference ($y\ll 1$), probability of
wrong flavour tagging ($w$) and finite accuracy in determination of $\Delta t$
(resolution function $R_{\rm sig}(\Delta t)$). Taking into account
these corrections, the final expected decay time distributions are
\begin{eqnarray}
\nonumber
&&\frac{d\Gamma(B^0\to f)}{d\Delta t}=e^{-\vert\Delta t\vert}\vert
    A_f\vert^2 \bigl[1+(1-2w)\cos(x\Delta t)\bigr]\otimes R_{\rm
    sig}(\Delta t)\\  
&&\frac{d\Gamma(\bar{B}^0\to f)}{d\Delta t}=e^{-\vert\Delta t\vert}\vert
    A_f\vert^2 \bigl[1-(1-2w)\cos(x\Delta t)\bigr]\otimes R_{\rm
    sig}(\Delta t)~~, 
\label{eq_4}
\end{eqnarray}
where the $\otimes$ sign denotes a convolution. The resolution function is composed as a convolution of several
Gaussian functions \cite{belle_vtxres}.  The
average accuracy of $\Delta t$ determination is around 1.4~ps (the 
lifetime of $B^0$ mesons is 1.53~ps \cite{pdg}).

\begin{figure}[h]
\centerline{\includegraphics[width=0.4\textwidth]{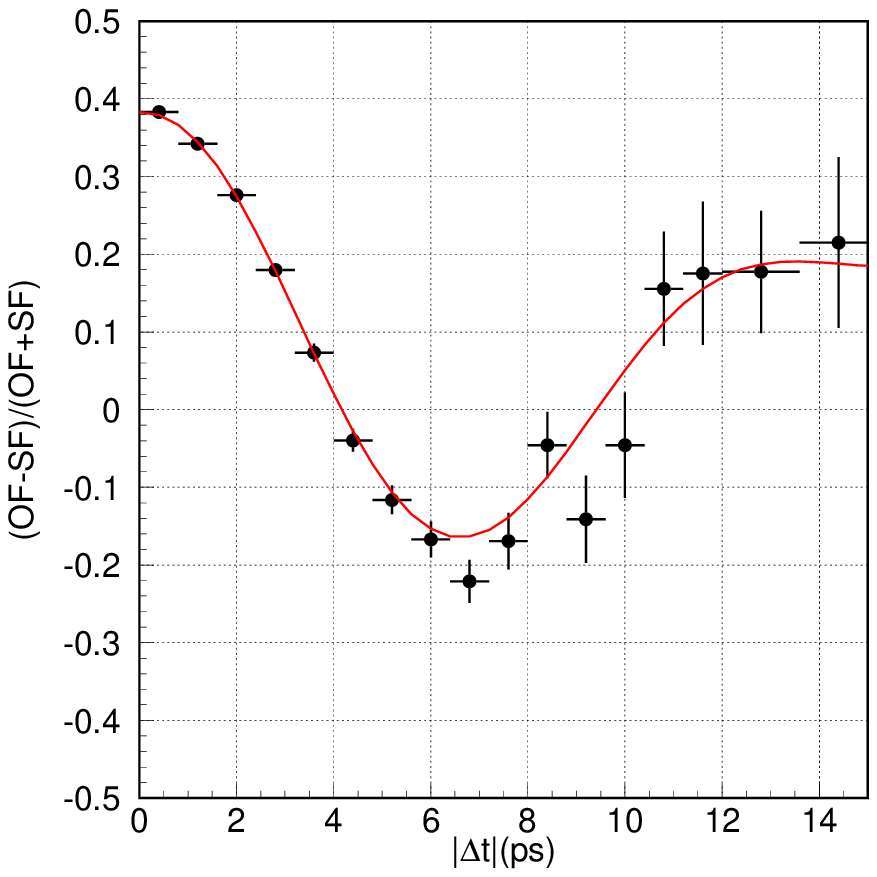}
\includegraphics[width=0.4\textwidth]{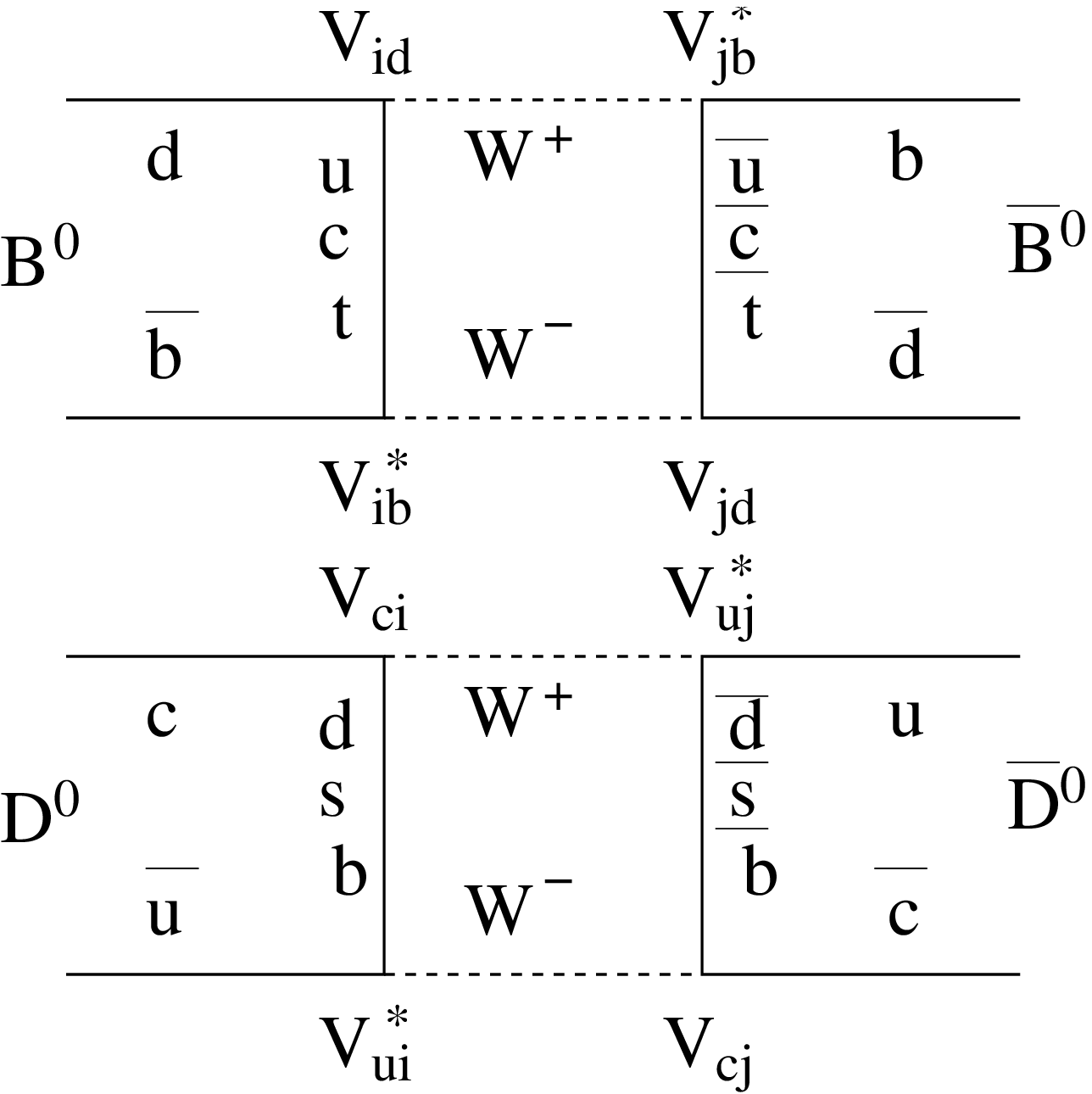}}
\caption{Left: Result of $B^0$ oscillation frequency measurement
  \cite{belle_dm} shown in the form of the asymmetry
  (\ref{eq_5}). Right: Loop diagram describing the $B^0\to\bar{B}^0$
transition (top) and the short distance contribution to $D^0\to\bar{D}^0$
  (bottom) transition.}
\label{fig_4}
\end{figure}

The most precise single measurement of $x$ \cite{belle_dm} uses several
flavour specific final states to reconstruct the signal $B^0$ meson
decays. Results are presented in Fig. \ref{fig_4} (left) in form of the asymmetry 
\begin{equation}
\frac{d\Gamma(B^0\to f)/d\Delta t-d\Gamma(\bar{B}^0\to f)/d\Delta t} 
{d\Gamma(B^0\to f)/d\Delta t+d\Gamma(\bar{B}^0\to f)/d\Delta
  t}=(1-2w)\cos{x\Delta t}\otimes R_{\rm
    sig}(\Delta t)~~. 
\label{eq_5}
\end{equation}
The average value of existing $\Delta m$ measurements \cite{hfag}, 
expressed in terms of $x=\Delta m/\bar{\Gamma}$, is 
$x=0.776\pm0.008$. 

\begin{figure}[h]
\centerline{\includegraphics[width=0.5\textwidth]{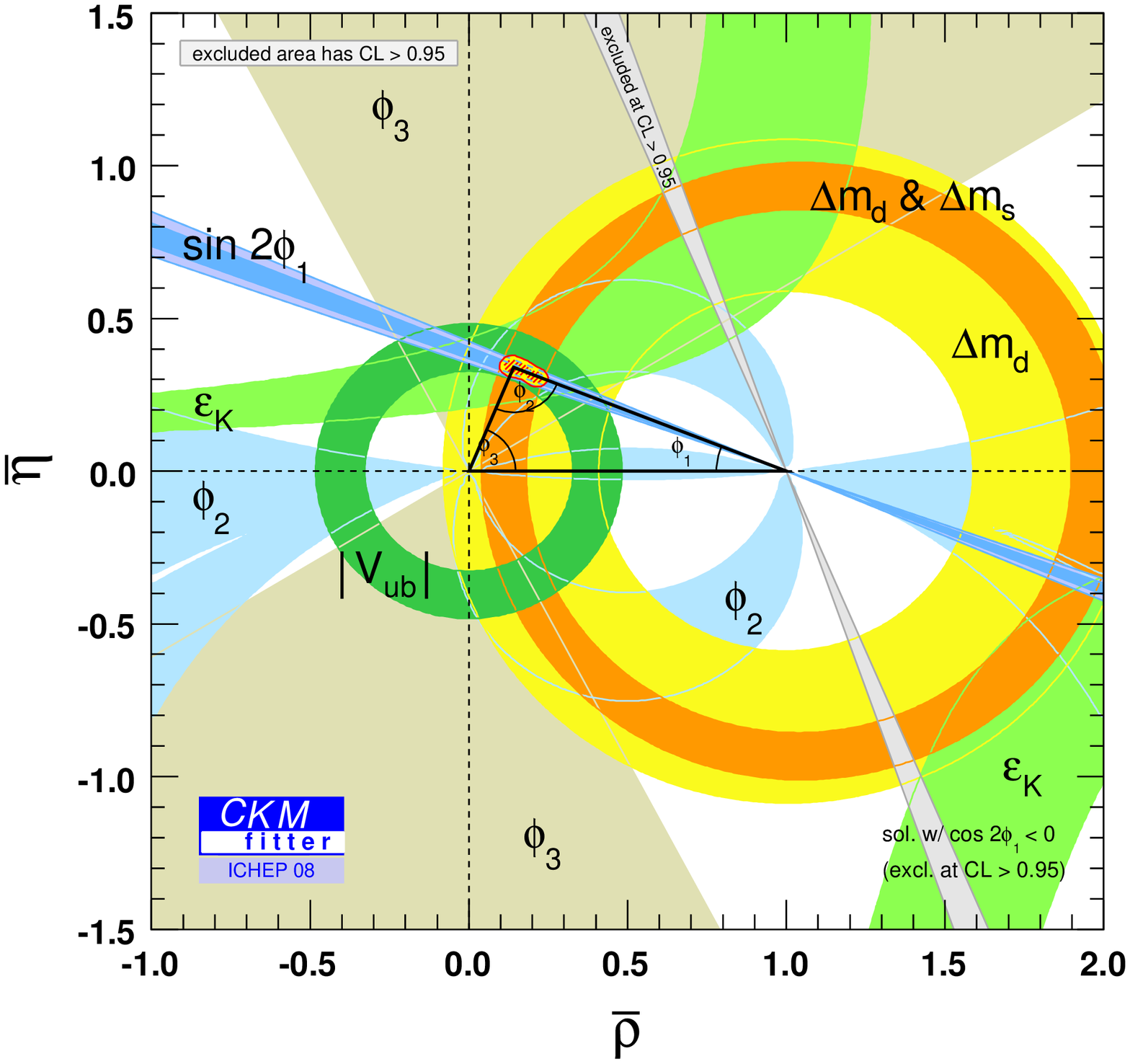}
\includegraphics[width=0.4\textwidth]{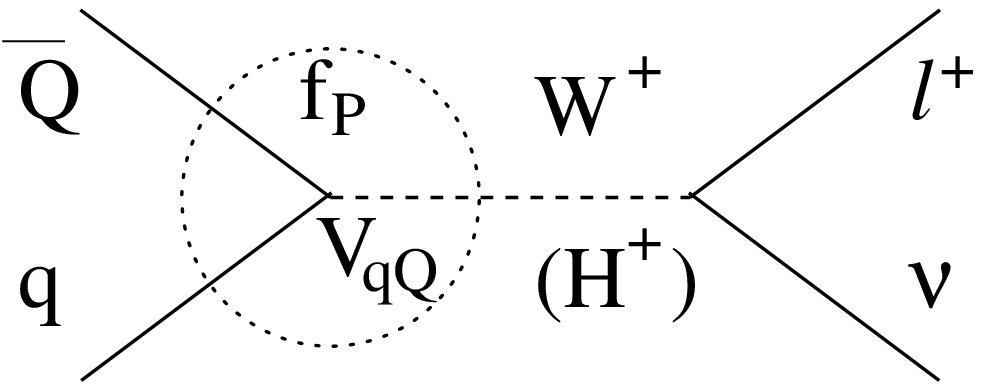}}
\caption{Left: Constraints in the $(\bar{\rho},\bar{\eta})$ plane arising
  from various measurement \cite{ckmfit}. The light shaded region
  denoted by $\Delta m_d$ represents the constraint
  from the $B^0$ oscillation frequency, and the dark shaded region
  denoted by $\Delta m_d \& \Delta m_s$ the constraint from the ratio
  of $B^0$ and $B_s^0$ oscillation frequencies. Right: 
Feynman diagram of a pseudoscalar meson leptonic decay.
  The QCD effects are described by the decay constant $f_P$. Beside
  the SM $W^+$ contribution also particles not included in the SM
  (like the charged Higgs boson) may contribute.}
\label{fig_5}
\end{figure}

Calculation of $\langle \bar{B}^0\vert
H_{eff}\vert B^0\rangle$ matrix element, visualized by the loop diagram of
Fig. \ref{fig_4} (right), results in \cite{buras_dm} 
\begin{equation}
\Delta
m_q=2\frac{G_F^2m_W^2\eta_Bm_{B_q}B_{B_q}f_{B_q}^2}{12\pi^2}S_0(m_t^2/m_W^2)|V_{tq}^\ast
V_{tb}|^2\bigl(1+{\cal{O}}(\frac{m_b^2}{m_t^2})\bigr)~.
\label{eq_6}
\end{equation} 
The equation is written using a subscript $q$ to
emphasize that the same relation is also appropriate for the system
of $B_s^0$ mesons. Using the measured value of the oscillation frequency
for $B^0$ mesons one can determine elements of the
Cabibbo-Kobayashi-Maskawa (CKM) matrix, if the QCD parameters
$\eta_B,~B_{B_q},~f_{B_q}$ are known\footnote{Function $S_0(x)$ is
  known.}. 
Due to their non-perturbative
nature these quantities are difficult to estimate (usualy the lattice
QCD calculations, LQCD, are exploited) and result in a large
uncertianty of CKM elements determination. The constraints from the measured value of
$\Delta m$ on parameters $(\bar{\rho},\bar{\eta})$ used to parametrize the CKM
matrix \cite{ckm_param} 
are shown in
Fig. \ref{fig_5} (left) \cite{ckmfit}. Since 2006 the oscillation frequency is measured
also in the system of $B_s^0$ mesons, $x_s=25.5\pm 0.6$
\cite{cdf_dm}. 
In the ratio of $\Delta m_s/\Delta m$ the QCD uncertainties cancel to a large
extent. The measured ratio $\Delta
m_s/\Delta m$ is thus much more constraining than $\Delta m$
constraint alone (Fig. \ref{fig_5} (left)), and actually at the moment
represents the most constraining measurement for the $\bar{\rho}$ among various flavour
physics studies. 

\subsection{Leptonic $B$ meson decays}
\label{sec_22}

Measurements of charged $B$ meson leptonic decays are interesting for
several reasons: theoretically they are easier to interpret compared to
semileptonic and hadronic decays, within the SM the measured rates can
potentially yield the value of the least known CKM element $V_{ub}$, and
they are sensitive to possible contributions of processes beyond the
SM. A Feynman diagram of an arbitrary pseudoscalar meson leptonic
decay is shown in Fig. \ref{fig_5} (right). The QCD effects are
described by a single parameter $f_P$, the meson decay constant
describing the overlap of the two quarks wave function.  
\begin{figure}[h]
\centerline{\includegraphics[width=0.4\textwidth]{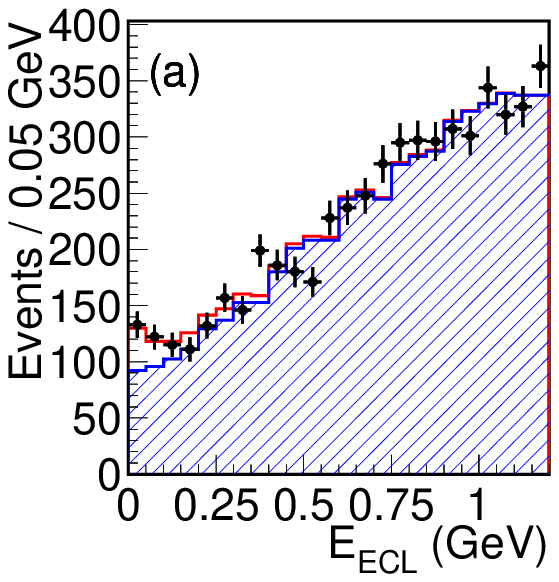}
\includegraphics[width=0.4\textwidth]{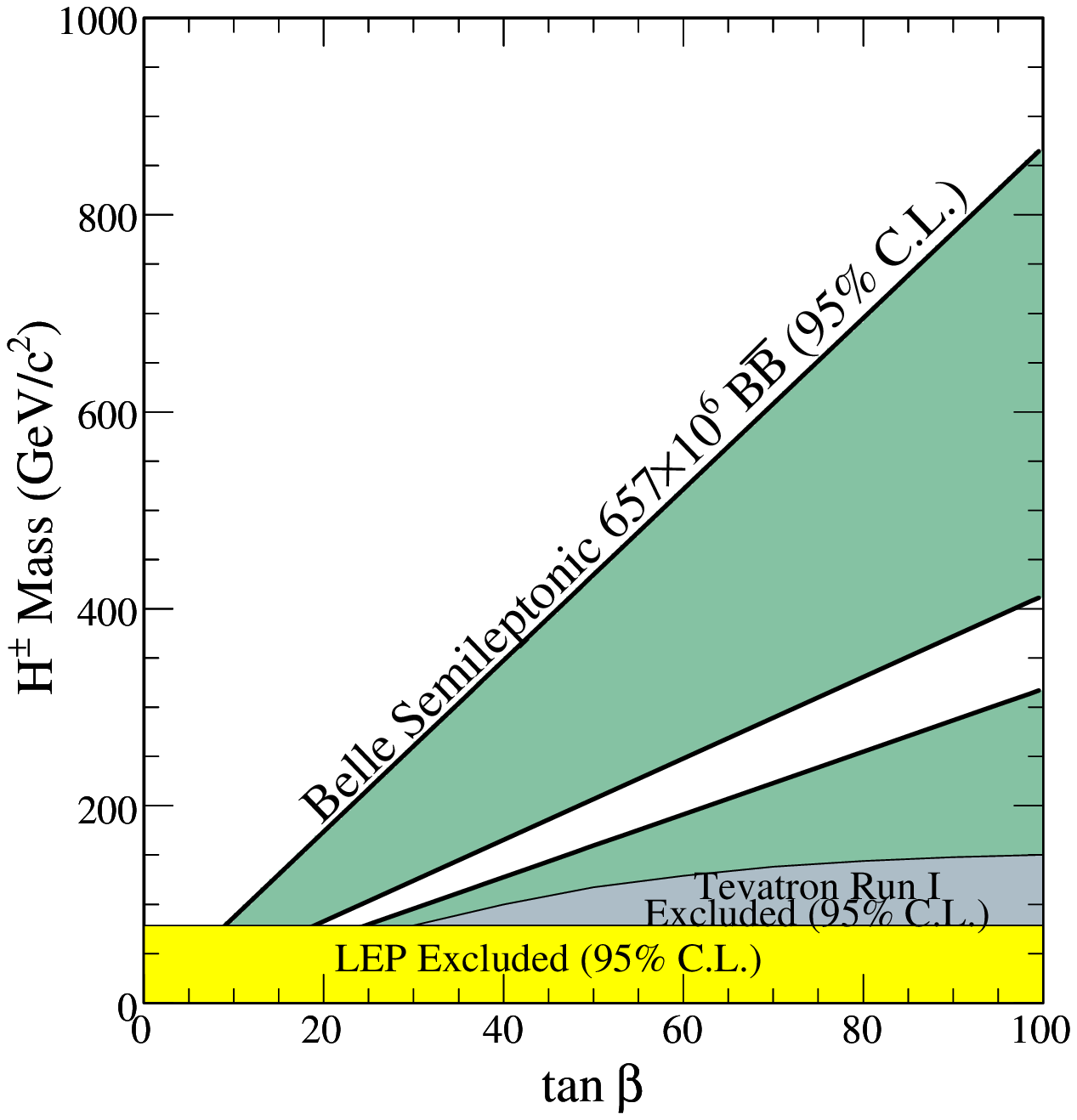}}
\caption{Left: Distribution
  of the energy in the electromagnetic calorimeter for selected $B^+\to\tau^+\nu_\tau$
  candidate events. The excess of data (points with error bars) above
  the expectation from simulation (histogram) is the signal. Right: 
Excluded region (shaded) in the $(m_H,\tan\beta )$ plane
arising from the measurement of $Br(B^+\to\tau^+\nu_\tau)$
\cite{belle_btaunu_semil}.}
\label{fig_6}
\end{figure}
The leptonic decays of a pseudoscalar mesons are helicity suppressed,
the expected ratios of decay widths are 
$1:4\times 10^{-3}:10^{-7}$ for the $\tau, \mu$ and $e$ 
decays, respectively. Despite the problems due to at least two undetected 
neutrinos in the final state the decays to $\tau$ leptons are the only
decays observed so far. 

The method of measurement consist of fully (partially) reconstructing the
accompanying $B$ meson using a large number of hadronic (semileptonic) decay
modes. After the particles assigned to the tagging meson are
successfully identified one searches for one or three charged tracks
originating from the $\tau$ decay. Finally the energy in the
electromagnetic calorimeter ($E_{\rm ECL}$) not assigned to the particles used in
the previous reconstruction is examined. Signal decays with only neutrinos
left in the final state are expected to peak at $E_{ECL}\sim 0$. The
$E_{\rm ECL}$ distribution of selected events in the measurement by
Belle \cite{belle_btaunu_semil} is shown in Fig. \ref{fig_6} (left). The
excess of events above the expectation from MC simulation at low values
of $E_{\rm ECL}$ is the signal for the $B^+\to\tau^+\nu_\tau$
decays. From the fit to the distribution the branching fraction 
$Br(B^+\to\tau^+\nu_\tau)=(1.65\pm
^{0.38}_{0.37}\pm^{0.35}_{0.37})\times 10^{-4}$ is obtained, where the main
contribution to the systematic error arises from the uncertainties in
the shape of the $E_{\rm ECL}$ signal and background distributions. A
similar measurement performed by BaBar \cite{babar_btaunu} yields 
$Br(B^+\to\tau^+\nu_\tau)=(1.2\pm 0.4\pm 0.4)\times 10^{-4}$, and
iclusion of the Belle measurement using the hadronic tagging
\cite{belle_btaunu_had} results in
the average of all measurements provided by Heavy Flavour Averaging
Group, $Br(B^+\to\tau^+\nu_\tau)=(1.51\pm 0.33)\times 10^{-4}$
\cite{hfag} \footnote{After the school an updated average of Belle and
  BaBar results appeared in
  \cite{sheldon}, $Br(B^+\to\tau^+\nu_\tau)=(1.73\pm 0.35)\times
  10^{-4}$.}. 

Calculation of $\Gamma(B^+\to\tau^+\nu_\tau)$ yields
\cite{pdg_deccons} 
\begin{equation}
\Gamma(B^+\to\tau^+\nu_\tau)=\frac{G_F^2}{8\pi}\vert
V_{ub}\vert^2f_B^2m_Bm_\tau^2 (1-\frac{m_\tau^2}{m_B^2})^2\bigl[ 
1-\frac{m_B^2}{m_H^2}\tan^2\beta\bigr]^2~~, 
\label{eq_7}
\end{equation}
where the last factor in brackets is a correction due to a possible contribution
of the charged Higgs boson. The measured value is in agreement with
the SM expectation (using LQCD prediction $f_B=(216\pm 22)$~MeV
\cite{lqcd_fb}, and $\vert V_{ub}\vert =(3.9\pm 0.5)\times 10^-3$
\cite{pdg_deccons}) and allows to put constraints on the parameters
$(m_H,\tan\beta)$ in the two Higgs doublet models ($m_H$ is the
charged Higgs boson mass and $\tan\beta$ is the
ratio of the vacuum expectation values). The constraints arising from
the Belle measurement are shown in Fig. \ref{fig_6} (right). 

\subsection{$b\to s\gamma$ decays}
\label{sec_23}

\begin{figure}[h]
\centerline{\includegraphics[width=0.3\textwidth]{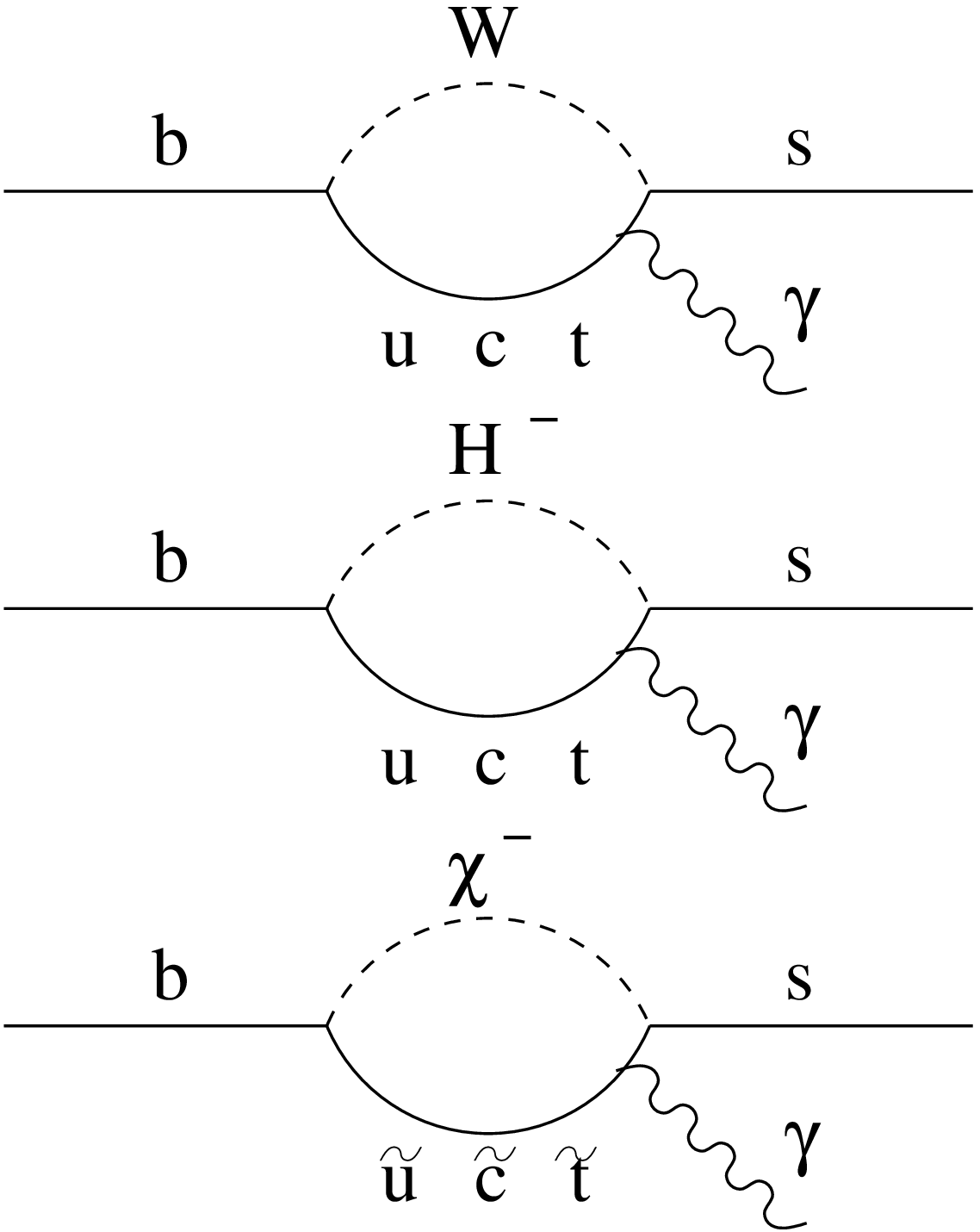}
\includegraphics[width=0.4\textwidth]{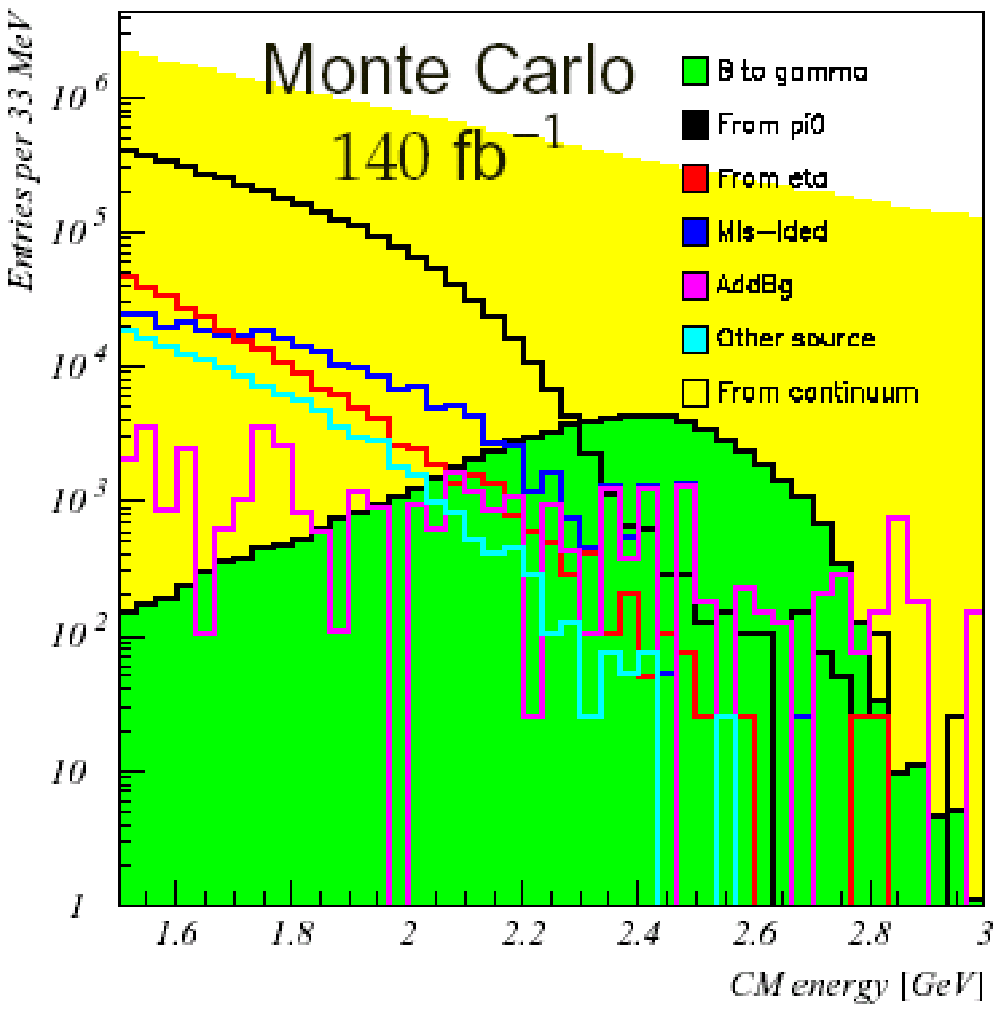}}
\caption{Left: SM (top) and some NP (middle and bottom) contributions 
to the $b\to s\gamma$ process. Right: Simulated photon energy
distributions from various processes. The smallest shaded region is
the contribution of  $b\to s\gamma$ (note the logarithmic scale).}
\label{fig_7}
\end{figure}

Decays involving the $b\to s\gamma$ transition cannot occur 
at the tree level in SM. Such a flavor changing neutral current (FCNC) 
is only 
possible as a higher order process and is thus sensitive to possible
contributions of New Physics (NP). Some possible diagrams, within and
beyond the SM, are shown in Fig. \ref{fig_7} (left). At the parton
level the photon energy in the CM frame is approximately half of the $b$ quark
mass. Also at the hadron level $E_\gamma$ is sensitive to $m_b$, which
is important for determination of $|V_{ub}|$ and $|V_{cb}|$ 
from semileptonic $B$ decays.  

There are both, theoretical and experimental difficulties in the
measurements. The former arise since in all experimental methods
there is a lower cut-off applied to $E_\gamma$. To determine the
branching fraction, for example, one has to extrapolate the partial
rate for $E_\gamma > E_{\rm cut}$ to the full energy region
using models, which introduces theoretical uncertainties. On the
experimental side the efforts are being made to lower the cut-off, but
this makes problems due to the huge backgrounds even more severe (see
Fig. \ref{fig_7} (right)). The name of the game is thus to suppress the
backgrounds to an acceptable level; {\it "Your background and
  environment is with you for life. No question about that."
  (S. Connery, 1930)}. 
Methods of reconstruction may be divided into inclusive, 
semi-inclusive and exclusive ones. In an inclusive measurement 
only the photon is reconstructed. From
the total $E_\gamma$ distribution of events recorded at the
$\Upsilon(4S)$ peak an analogous distribution of events, recorded
60~MeV below the peak is subtracted. The latter represents only the photons
arising from $e^+e^-\to q\bar{q}$ (continuum) events, and if the distribution is
scaled according to the integrated luminosity of both samples, the
remainder after the subtraction represents the energy distribution of photons from $B$ meson
decays. A search is made for photon pairs consistent with $\pi^0$ or
$\eta$ decays and such $\gamma$'s are removed from the selected
sample. The remaining background is estimated using
simulated samples but normalized using data control samples. 

\begin{figure}[h]
\centerline{\includegraphics[width=0.3\textwidth]{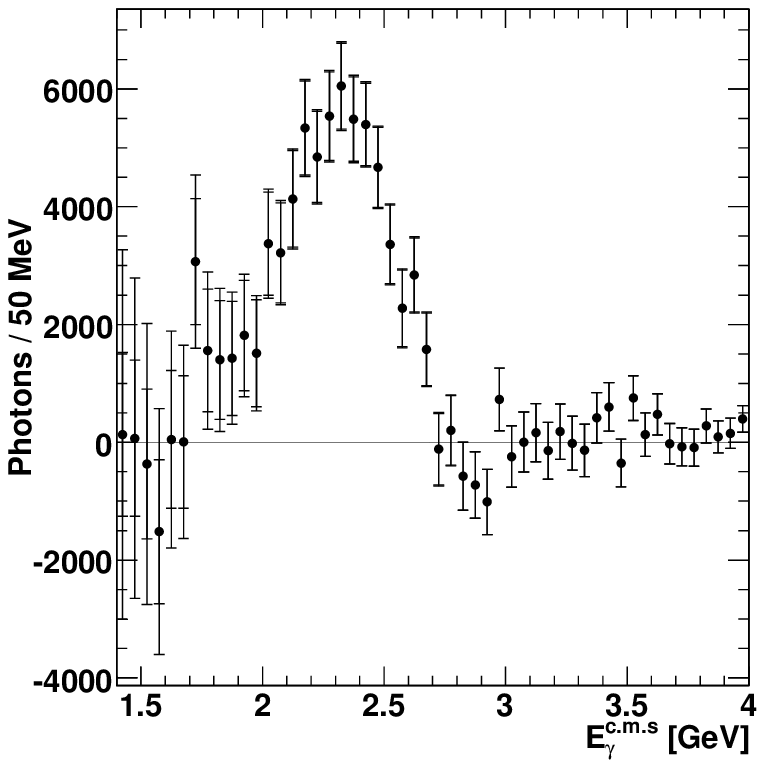}
\includegraphics[width=0.4\textwidth]{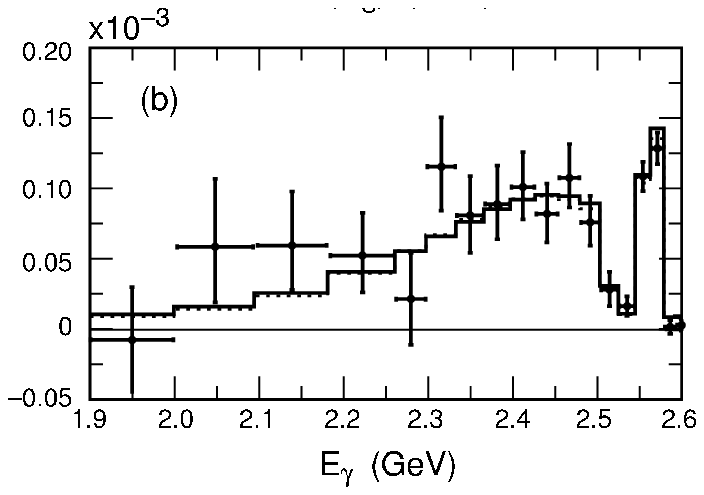}}
\caption{Left: Raw CM system photon energy distribution for inclusively
  reconstructed $b\to s\gamma$ decays \cite{belle_bsg}. Right:
  Differential branching fraction of $B\to X_s\gamma$ as a function of $E_\gamma$
  obtained in semi-inclusive measurement \cite{babar_bsg}.}
\label{fig_8}
\end{figure}

Result of such an inclusive method is shown in Fig. \ref{fig_8} (left)
\cite{belle_bsg}. 
$E_\gamma$ distribution peaks at around half of the $b$ quark mass and is
consistent with zero above the kinematic limit for $B\to K\gamma$
decays, confirming the correctness of the subtraction procedure. To
determine the branching fraction and the correct shape of the energy
distribution one has to apply a deconvolution method to the raw
spectrum, correct it for the efficiency of reconstruction, subtract a
simulated contribution of $b\to d \gamma$ decays ($\sim$4\%) and make
the transformation to the $B$ meson rest frame. The partial branching
fraction in the interval $E_\gamma > 1.7$~GeV is found to be 
$Br(b\to s\gamma)=(3.31\pm 0.19\pm 0.37\pm 0.01)\times 10^{-4}$. The
last uncertainty is due to the boost from the CM to the $B$ meson
frame, and the largest systematic uncertainty arises from the
normalization of backgrounds other than $\pi^0$ and $\eta$. 

As an example of a semi-inclusive measurement we present the analysis
of $B\to X_s\gamma$ by BaBar collaboration \cite{babar_bsg}, 
where $X_s$ represent a sum of various decay modes
with $K^\pm,~K_S,~\pi^\pm$ and $\eta$ mesons in the final state. The photon
energy is in this method calculated from the invariant mass of the
hadronic system $m(X_s)$ which results in a better resolution compared to
the measured photon energy in the electromagnetic calorimeter. The background
is suppressed using neural network for the rejection of continuum events
and vetoes for $\gamma$'s from $\pi^0$ and $\eta$ mesons. To calculate
the branching fraction for $B\to X_s\gamma$ the number of observed
events must be corrected for the fraction of decays not taken into
account in the reconstruction (25\% at low $m(X_s)$ 
due a to non-inclusion of $K_L$, and higher at higher masses). The resulting
differential branching fraction for $E_\gamma > 1.9$~GeV is shown in
Fig. \ref{fig_8} (right). The integral rate in the $E_\gamma > 1.9$~GeV
interval is found to be 
$Br(b\to s\gamma)=(3.27\pm 0.18\pm ^{0.55}_{0.40}\pm
^{0.04}_{0.12})\times 10^{-4}$, where the last error is due to the QCD
parameters affecting the efficiency. 

The measured branching fractions impose limits on possible
contribution of charged Higgs boson. The world average of inclusive
branching fraction is $Br(b\to s\gamma)=(3.52\pm 0.23\pm 0.09)\times
10^{-4}$ \cite{hfag}. The 95\% C.L. limit following from
\cite{misiak_bsg} is $m_H>300$~GeV for any value of $\tan\beta$. In
all measurements also the first and the second moment of the photon energy spectra are
determined. These can be expressed with the same QCD parameters entering
also the determination of $|V_{ub}|$ and $|V_{cb}|$ in inclusive semileptonic $B$
decays. Details of a simultaneous fit performed to photon energy
spectrum in $b\to s\gamma$ and lepton momentum and hadronic mass
spectra in semileptonic decays
to determine the elements of CKM matrix is described in \cite{hfag}. 

\subsection{$b\to s{\cal{\ell^+\ell^-}}$ decays}
\label{sec_24}

\begin{figure}[h]
\centerline{\includegraphics[width=0.3\textwidth]{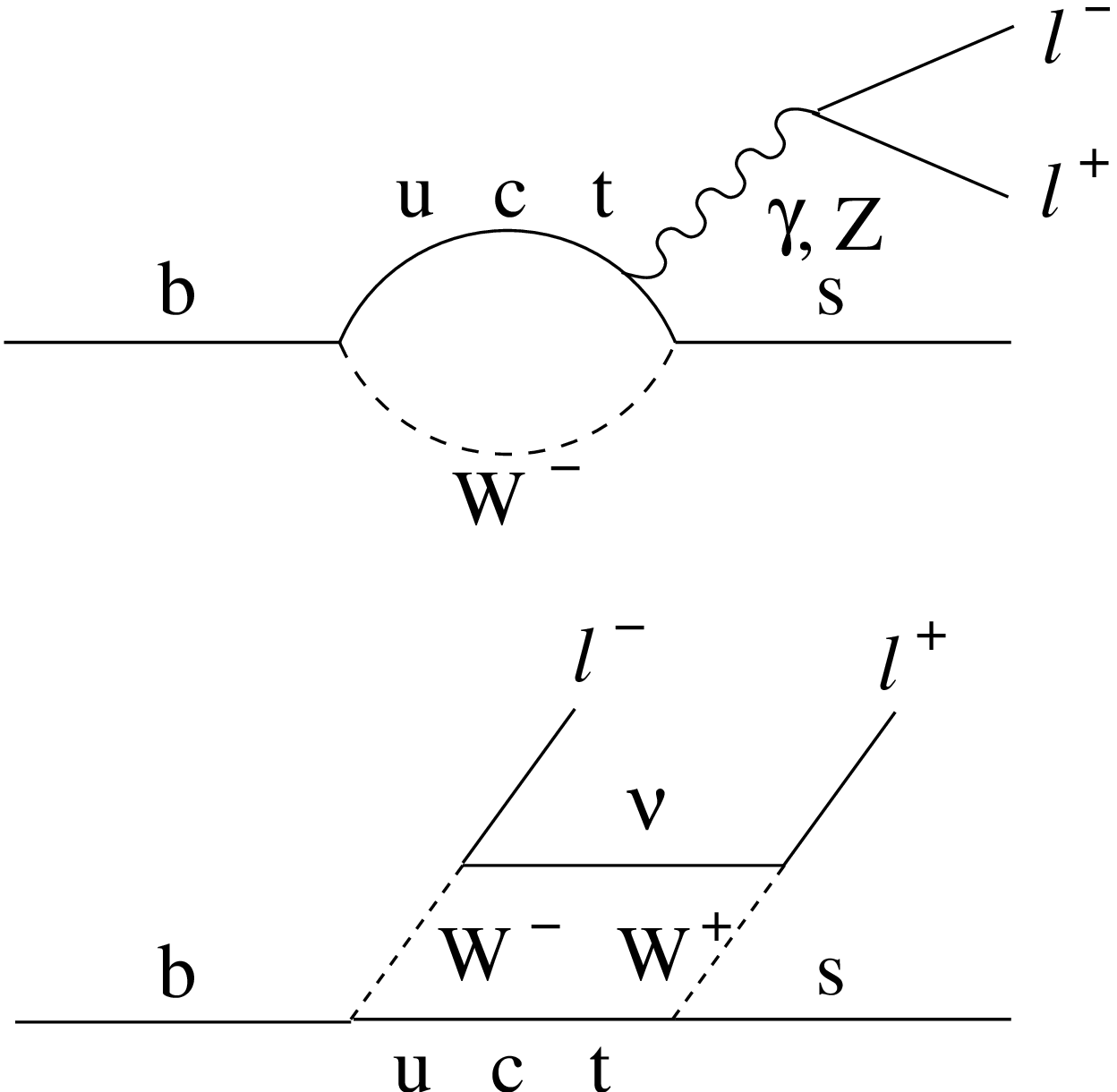}
\includegraphics[width=0.4\textwidth]{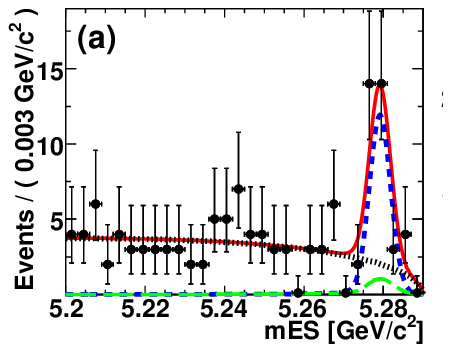}}
\caption{Left: SM Feynman diagrams for $b\to s{\cal{\ell^+\ell^-}}$. Right:
  Energy substituted $B$ meson mass for reconstructed $B\to K^\ast{\cal{\ell^+\ell^-}}$ decays. The
  distribution is shown for a part of the full sample with low $q^2$
  \cite{babar_bsll}.}
\label{fig_9}
\end{figure}

Decays involving the $b\to s{\cal{\ell^+\ell^-}}$ parton process are
  another example of a FCNC transition. From that point of view they
  are interesting for the same reasons as the $b\to s\gamma$
  decays. Again, the inclusive decays are theoretically easier to
  interpret than the exclusive ones. Nevertheless, a lot of work has
  been done in identifying the observables in exclusive decays,
  especially $B\to K^\ast{\cal{\ell^+\ell^-}}$, for which the
  theoretical uncertainties are small \cite{hurth_bsll}. Feynman
  diagrams contributing to $b\to s{\cal{\ell^+\ell^-}}$ are shown in
  Fig. \ref{fig_9} (left). 
The differential decay rate $d\Gamma/dq^2$,
  where $q^2$ is the invariant mass of the lepton pair, 
  can
  be described in terms of effective Wilson coefficients $C_7^{\rm
  eff},~C_9^{\rm eff}$ and $C_{10}^{\rm eff}$, 
  which include the perturbative part of the process and thus
  dependence on heavy masses of SM particles $m_W,~m_t$, as well as
  on possible NP masses $m_{\rm NP}$. The absolute value of the coefficient
  $C_7^{\rm eff}$ can be constrained from the measured rate of $b\to s\gamma$
  process, while in $b\to s{\cal{\ell^+\ell^-}}$ additional information
  (on sign of $C_7^{\rm eff}$, as well as $C_9^{\rm eff},~C_{10}^{\rm
  eff}$) can be obtained due to
  the interference of the two amplitudes shown in Fig. \ref{fig_9} (left). NP could change the values
  of the Wilson coefficients as well as add new operators causing the
  transition. In exclusive $B\to K^\ast{\cal{\ell^+\ell^-}}$ decays the
  theoretical description includes beside
  the Wilson coefficients also the non-perturbative part expressed
  by the form factors which are predicted with an accuracy of around
  30\% \cite{ali_bsll}. This uncertainty is significantly reduced
  in some observables arising from the study of angular
  distributions, like the lepton forward-backward asymmetry ($A_{FB}$) and
  the fraction of longitudinally polarized $K^\ast$'s ($F_L$). 

BaBar performed a study of $B\to K^\ast{\cal{\ell^+\ell^-}}$ decays
in \cite{babar_bsll}. The reconstruction proceeds through
identification of a lepton pair ($\mu^+\mu^-$ or $e^+e^-$) with
invariant mass not in the range of charmonium states $J/\psi$ or
$\psi(2S)$. The $K^\ast$ can be either charged or neutral,
reconstructed through $K\pi,~K\pi^0$ and $K_S\pi$ final states. The
signal can be seen in the energy substituted $B$ meson mass,
$M_{ES}=(E_{CM}/2)^2-(\sum_i\vec{p}_i)^2$, where 
$\vec{p}_i$ and $E_{CM}$ are the $B$ decay products momenta and
$e^+e^-$ collision energy, respectively, calculated in the CM
frame (Fig. \ref{fig_9} (right)). The background is composed of
combinatorial one (described by reconstructing events with $\mu^\pm
e^\mp$ lepton pairs), hadrons misidentified as muons (in
$\mu^\pm\mu^\mp$ channel) and peaking background from $B\to D\pi$
decays where the charmed meson decays to $K^\ast\pi$ (vetoed by
requiring the invariant mass of $K^\ast\pi$ not to be consistent with a $D$
meson). 

\begin{figure}[h]
\centerline{\includegraphics[width=0.4\textwidth]{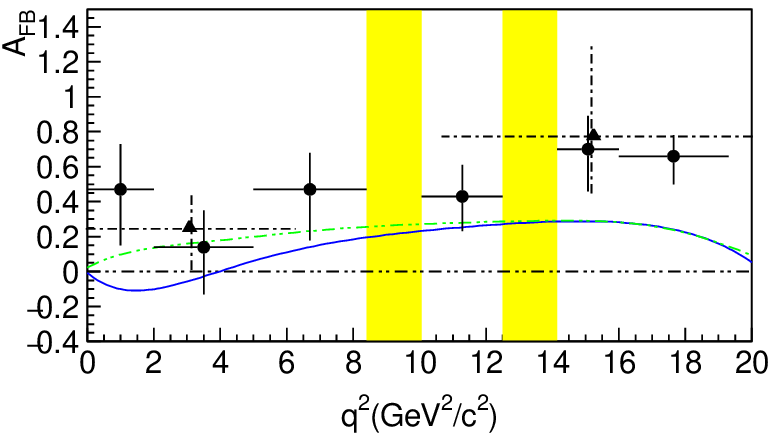}
\includegraphics[width=0.6\textwidth]{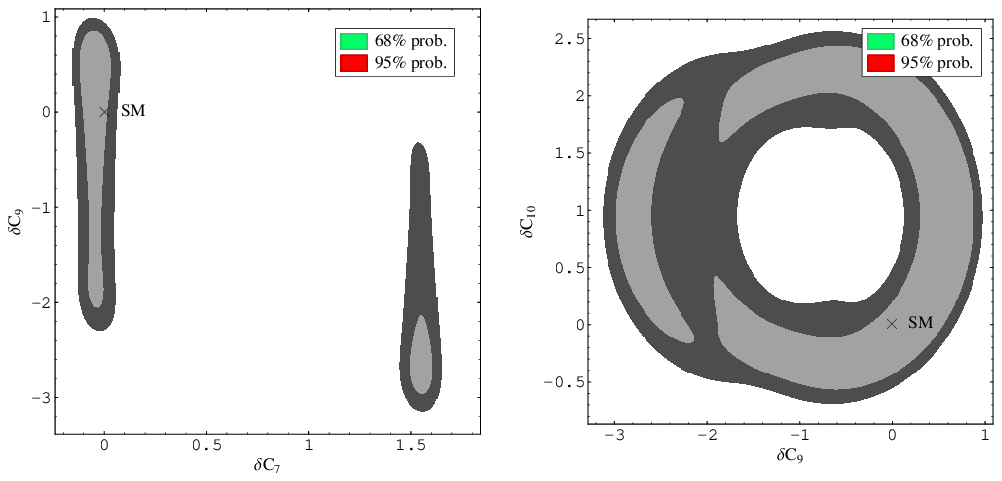}}
\caption{Left: Measured $A_{FB}$ as a function of $q^2$ in $B\to
  K^\ast{\cal{\ell^+\ell^-}}$ decays. The data points marked with
  triangles (dashed-dotted error bars) are from \cite{babar_bsll} and
  circular data points (full error bars) from \cite{belle_bsll}. The
  shaded regions represent $q^2$ intervals not included in the
  measurement ($J/\psi,~\psi(2S)$ regions). Full line represents the
  SM prediction and the dashed curve prediction with $C_7^{\rm
  eff}=-C_7^{\rm SM,~eff}$. Right: Constraints on NP contributions to
  Wilson coefficients arising from measurements of various FCNC
  processes \cite{kamenik}. The SM corresponds to $\delta
  C_i=0$. Light shaded areas represent the 68\% C.L. and dark shaded
  the 95\% C.L. region.}
\label{fig_10}
\end{figure}

For the reconstructed events the distribution of the kaon helicity
angle in the rest frame of $K^\ast$ is investigated to obtain the
fraction of longitudinally polarized $K^\ast$'s ($F_L$). $F_L$ value is
then used as an input to the fit to the distribution of the angle between the
lepton and $K^\ast$ in the ${\cal{\ell^+\ell^-}}$ rest frame. This
  angle follows a
  $1+3F_L+(1-F_L)\cos^2{\theta_{\cal{\ell}}}+(8/3)A_{FB}\cos{\theta_{\cal{\ell}}}$ 
distribution. The $A_{FB}$ is the lepton forward-backward asymmetry
which can be predicted in terms of the Wilson coefficients. In
Fig. \ref{fig_10} (left) measured $A_{FB}$ is shown as a function of
$q^2$ for the measurement by BaBar as well as the most recent
measurement by Belle collaboration \cite{belle_bsll}. The measured
values are compared to the SM prediction and the expectation for the
Wilson coefficient $C_7^{\rm eff}$ of reversed sign. In general the
measurements seems to be shifted to larger asymmetry values than
predicted. 

Similarly as for the $b\to s\gamma$ decays, also for $b\to
s{\cal{\ell^+\ell^-}}$ semi-inclusive measurements have been performed
by summing up various hadronic decay modes of the strange quark system
($K_S$ or $K^\pm$ with 0 to 4 pions) accounting for around 70\% of the
total decay rate. The average of the branching fraction measurements
is $Br(B\to X_s{\cal{\ell^+\ell^-}})=(4.50\pm ^{1.03}_{1.01})\times
10^{-6}$ \cite{hfag}. 

Various measurements of FCNC can be combined to put constraints on
possible NP contribution to Wilson coefficients. Within a Minimal
Flavour Violation scenario these constraints are presented in
Fig. \ref{fig_10} (right) \cite{kamenik}. Measurements of $Br(B\to
X_s\gamma)$, $Br(B\to X_s{\cal{\ell^+\ell^-}})$, $Br(K\to\pi\nu\nu)$
and $Br(B_s\to\mu\mu)$ are used as the input. The combination of
measurements is consistent with the SM ($\delta C_i=0$) although
there are large areas corresponding to non-SM contributions possible.

\section{Lecture II}
{\it "{\bf Charm} is...a way of getting the answer yes without having to ask
  any clear question." (A. Camus, 1913-1960)}
\subsection{$D$ meson oscillations}
\label{sec_31}

The second lecture is devoted to results in physics
of charmed hadrons from $B$-factories. Charm physics in recent years gained in interest
of both, experimental and theoretical physicists, mainly due to new
interesting results from Belle and BaBar. Both experiments are not
only factories of $B$ mesons but also of charmed hadrons (see Section
\ref{sec_1} and Fig. \ref{fig_1}). Contemporary charm physics has a
twofold impact: as a ground of theory predictions tests, mainly tests
of LQCD, and as a self-standing field of SM measurements and NP
searches. An example of the first kind are the measurements of charmed
meson decay constants, to be compared to LQCD calculations, to verify
those and thus enable a more reliable estimates of the CKM matrix elements
from the measurements in the $B$ meson sector. The outstanding examples of the
second group of measurements are recent observations of $D^0$ mixing
and searches for the $CP$ violation in processes involving charmed
hadrons. 

Neutral $D$ mesons are the only neutral meson system composed of
up-like quarks. Hence a different contribution of virtual new particles
than in the mixing of other neutral mesons is possible in the loops of diagrams describing the
$D^0\leftrightarrow \bar{D}^0$ transition 
(Fig. \ref{fig_4} (right)). However, the short distance contribution to
the mixing rate, illustrated by the box diagram, is extremely
small. The reason is the effective GIM suppression; calculation of
the amplitude for this transition reveals \cite{burdman} that it is proportional to 
$V_{cs}^\ast V_{cd}^\ast V_{ud}V_{us}(m_s^2-m_d^2)/m_c^2$. Hence the
amplitude is doubly Cabibbo suppressed, and furthermore arises only as
a consequence of SU(3) flavour symmetry breaking. The
resulting oscillation frequency defined in Sect. \ref{sec_21} is
$|x_D|={\cal{O}}(10^{-5})$. This unobservable effect is hindered by long
distance contribution to the transition amplitude, for example from
states accessible to both, $D^0$ and $\bar{D}^0$ (e.g. $D^0\to
K^+K^-\to\bar{D}^0$). This contribution is difficult to
estimate. Current calculations \cite{dmix_th} within the SM predict
$|x_D|,~|y_D|\lesssim{\cal{O}}(10^{-2})$. The result illustrates the order of
magnitude of the mixing parameters to be expected in the $D^0$ system
(compare to measured values of $x,~x_s$ given in Sect. \ref{sec_21})
as well as the large theoretical uncertainty of the predictions. 

The time evolution of an initially produced $D^0$ meson follows
Eq. (\ref{eq_2}) with the simplification due to $|x_D|,~|y_D|\ll 1$: 
\begin{equation}
\frac{d\Gamma(D^0\to f)}{dt}=e^{-t}\vert A_f
+\frac{q}{p}\frac{ix_D+y_D}{2}A_{\bar{f}}\vert^2+{\cal{O}}(x_D^3,~y_D^3)~~.
\label{eq_8}
\end{equation}
The time integrated rate for an initially produced $D^0$ meson to
decay as a $\bar{D}^0$, 
$R_M=(x_D^2+y_D^2)/2\sim 10^{-4}$, is small and represents the reason for
a 31 years time span between the discovery of $D^0$ mesons and
the experimental observation of the $D^0$ mixing. 
%According to the famous
%writer, {\it "The duration of passion is proportionate with the
%  original resistance of the woman." (H. de Balzac, 1799 - 1850)},
%this was a further motivation to perform such a measurement. 

There are several methods and selection criteria common to various
measurements of $D^0$ mixing. Tagging of the flavour of an initially produced $D^0$
meson is achieved by reconstruction of
decays $D^{\ast +}\to D^0\pi_s^+$ or $D^{\ast -}\to
\bar{D}^0\pi_s^-$. The charge of the characteristic low momentum
pion $\pi_s$ determines the tag. The
energy released in the $D^\ast$ decay, $q=m(D^\ast)-m(D^0)-m_\pi~~$, 
has a narrow peak for the signal events and thus
helps in rejecting the combinatorial background. $D^0$ 
mesons produced in $B$ decays have a different decay length distribution
and kinematic properties than the mesons 
produced in fragmentation. In order to obtain a sample of neutral
mesons with uniform properties one selects $D^\ast$ mesons with
momentum above the kinematic limit for the $B$ meson decays. The decay
time is obtained from the reconstructed momentum and decay length of
$D^0$ meson, and the latter is obtained from a common vertex of $D^0$ 
decay products and an intersection point of $D^0$ momentum vector and
the $e^+e^-$ interaction region. 

Methods of measuring the mixing parameters as well as sensitivities
depend on specific final states chosen. The first to be described are
decays to $CP$ eigenstate $f_{CP}$.  In the limit
of negligible $CP$ symmetry violation ($CPV$, described in Section
\ref{sec_32}) 
the mass eigenstates $D_{1,2}$ coincide with the 
$CP$ eigenstates (in case of no $CPV$ $q/p=1$, see Eq. (\ref{eq_1})). 
In decays $D^0\to f_{CP}$ only the mass eigenstate 
component of $D^0$ with the $CP$ eigenvalue equal to the one of
$f_{CP}$ contributes. 
By measuring the lifetime of $D^0$ in decays
to $f_{CP}$ one thus determines the corresponding $1/\Gamma_1$ or
$1/\Gamma_2$. On the other
hand, flavour specific final states like $K^-\pi^+$ have a mixed $CP$
symmetry. The measured value of the effective lifetime in these decays 
corresponds to a mixture of $1/\Gamma_1$ and 
$1/\Gamma_2$. The relation between the two lifetimes can be written as 
\cite{bergman}
\begin{equation}
\tau(f_{CP})=\frac{\tau(D^0)}{1+\eta_f y_{CP}}~~, 
\label{eq_9}
\end{equation}
where $\tau(f_{CP})$ and $\tau(D^0)$ are the lifetimes measured in
$D^0\to f_{CP}$ and $D^0\to K^-\pi^+$, respectively. 
$\eta_f=\pm 1$ denotes the $CP$ eigenvalue of $f_{CP}$. 
The relative difference of
the lifetimes is described by the parameter $y_{CP}$. Expressed in terms of the mixing parameters, $y_{CP}$ reads
\cite{bergman} $y_{CP}=y_D\cos{\phi}-(1/2)A_Mx_D\sin{\phi}$,  
with $A_M$ and $\phi$ describing the $CPV$ in mixing and in interference
between mixing and decays, respectively. In case of no $CPV$, $A_M=\phi
=0$ and $y_{CP}=y_D$. 

\begin{figure}[h]
\centerline{\includegraphics[width=0.6\textwidth]{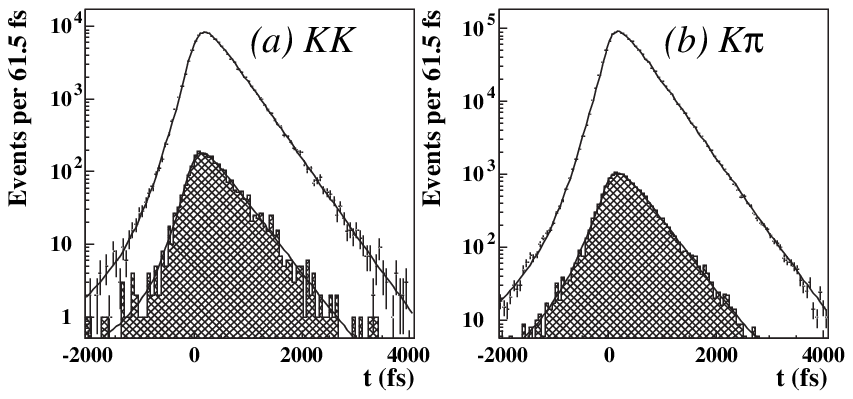}
\includegraphics[width=0.31\textwidth]{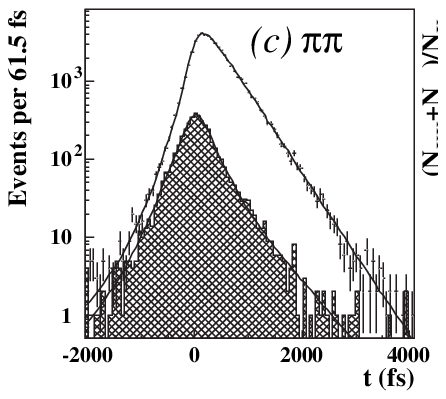}}
\caption{Decay time distributions of $D^0\to h^+h^-$ ($h=K,~\pi$)
  \cite{belle_ycp}. 
Hatched histogram is the
  contribution of background obtained from $m(h^+h^-)$ sidebands. Full line is the result of a
  simultaneous fit to all three distributions with $y_{CP}$ as a
  common free parameter.}
\label{fig_11}
\end{figure}

The measurement of $y_{CP}$ by Belle \cite{belle_ycp} represents the first
evidence of $D^0$ mixing \footnote{Published simultaneously with
  the measurement of $D^0\to K^+\pi^-$ decays by BaBar \cite{babar_kpi}
  which also gives evidence of the mixing.}. Number of reconstructed
decays to $CP$-even states $K^+K^-$ and $\pi^+\pi^-$ were $110\times
10^3$ and $50\times 10^3$, with purities of $98\%$ and $92\%$,
respectively. A simultaneous fit to the decay time distributions of
$KK,~\pi\pi$ and $K\pi$ decays was performed with $y_{CP}$ as a common
free parameter. In order to perform a precision measurement of
lifetime in each of the decay modes a special care should be devoted
to a proper description of $t$ resolution function in various
data-taking periods (for
details the reader is referred to the original publication). The $t$
distributions and the result of the fit are presented in
Fig. \ref{fig_11}. The quality of the fit ($\chi^2/n.d.f.=313/289$)
confirms an accurate description of the resolution effects. 
The measured value of $y_{CP}$ is $(1.31\pm 0.32\pm 0.25)\%$ and the
largest contribution to the systematic uncertainty arises due to a
possible small detector induced bias in the decay time
determination. $y_{CP}$ 
deviates from the null value by more than three standard deviations
including the systematic uncertainty. This evidence is confirmed by a
similar measurement performed by the BaBar collaboration
\cite{babar_ycp}, 
finding $y_{CP}=(1.24\pm 0.30\pm 0.13)\%$. 

\begin{figure}[h]
\centerline{\includegraphics[width=0.4\textwidth]{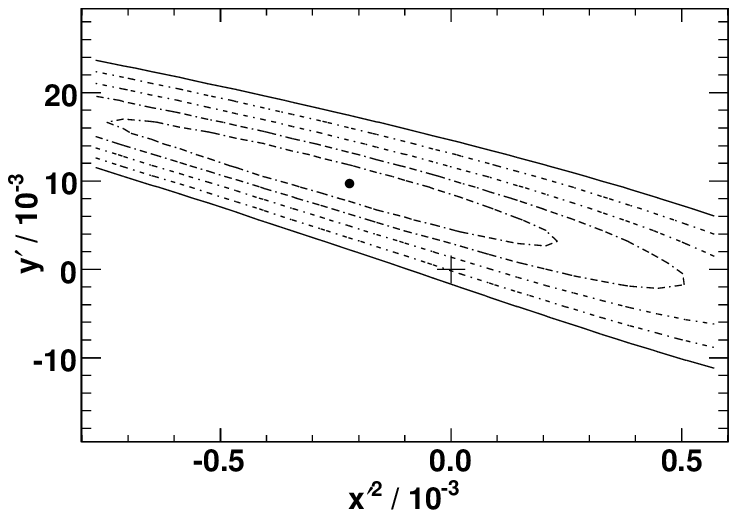}
\includegraphics[width=0.4\textwidth]{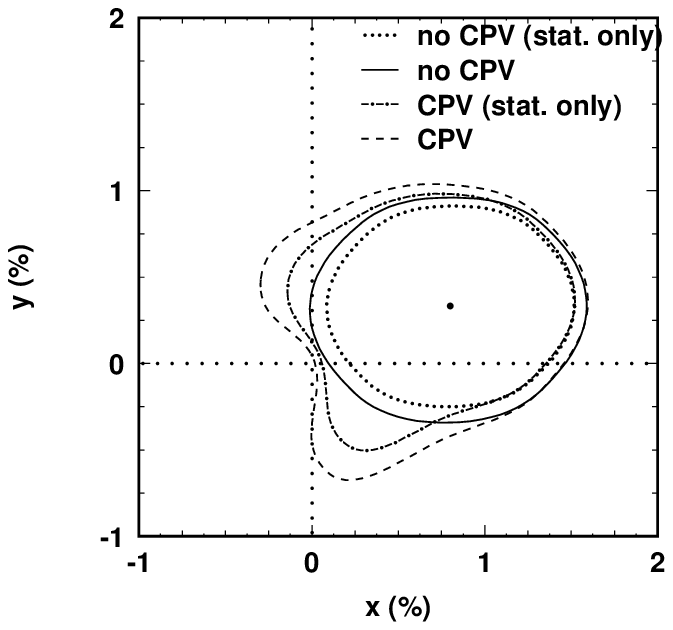}}
\caption{Left: Likelihood contours in $(x^{\prime 2},y^\prime)$ plane
  arising from the measurement of $D^0\to K^+\pi^-$ decays
  \cite{babar_kpi}. The lines correspond to 1-5 $\sigma$ C.L., the
  dot represents the central value and the cross the no-mixing
  point. Right: 95\% C.L. contour of $(x_D,y_D)$ as determined in
  the time-dependent Dalitz study of $D^0\to K_S\pi^+\pi^-$
  \cite{belle_kspipi}. Regions arising from the fit with neglected and
  allowed $CPV$, as well as neglecting or incorporating the systematic
  uncertainty are shown.}
\label{fig_12}
\end{figure}

Another possibility to look for the effect of mixing represent decays of
initially produced $D^0$'s to a wrong-sign final state $K^+\pi^-$. While
the more abundant $D^0$ decays lead to the $K^-\pi^+$ charge
combination, the wrong-sign combination can be reached through doubly
Cabibbo suppressed (DCS) decays or through a $D^0\to\bar{D}^0$ mixing
followed by a Cabibbo favored (CF) $\bar{D}^0$ decay. In order to separate
the mixing contribution from the DCS decays an analysis of the decay
time distribution must be performed. The $t$-dependent decay rate, 
$d\Gamma(D^0\to K^+\pi^-)\propto [R_D+\sqrt{R_D}y_D^\prime t 
  +(1/4)(x_D^{\prime 2}+y_D^{\prime 2})t^2]e^{-t}$, 
consists of three terms corresponding to DCS term ($R_D$), mixing term
($x_D^{\prime 2}+y_D^{\prime 2}$) and the
interference between the two. Additional complication in the
interpretation of the result arises since the decay rate depends on
parameters $x^\prime$ and $y^\prime$ which are the mixing parameters
rotated by a strong phase difference between the amplitudes of CF and
DCS decays. In \cite{babar_kpi} BaBar collaboration
fitted the $t$ distribution of around 4000 reconstructed wrong-sign
decays. Result of the fit is presented in Fig. \ref{fig_12} (left) in
terms of the allowed region in $(x^{\prime 2},y^\prime)$ plane. While 
the central value is in the physically forbidden region ($x^{\prime 2}<0$) the
no-mixing point ($(x^{\prime 2},y^\prime)=(0,0)$) is excluded by a 
confidence level corresponding to 3.9 standard deviations. 
Numerically they find $x^{\prime 2}=(-0.22\pm 0.33\pm 0.21)\times
10^{-3}$ and $y^\prime=(0.97\pm 0.44\pm 0.31)\%$. 

The method which allows for a direct determination of both mixing
parameters, $x_D$ and $y_D$, is the study of decays into self
conjugated multi-body final states. Several intermediate resonances
can contribute to such a final state. In the recent measurement by 
Belle \cite{belle_kspipi} the $K_S\pi^+\pi^-$ final state was
analyzed, where contributions from CF decays (e.g. $D^0\to K^{\ast
  -}\pi^+$), DCS decays (e.g. $D^0\to K^{\ast
  +}\pi^-$) and decays to $CP$ eigenstates (e.g. $D^0\to \rho^0 K_S$)
are present. Individual contributions can be identified by analyzing
the Dalitz distribution of the decay. Due to the interference among
different 
types of decays it is possible to
determine their relative phases (unlike in $D^0\to K^+\pi^-$ decays
where the relative phase between DCS and CF decays cannot be determined). 
And most importantly, since these types of intermediate states also exhibit 
a specific time evolution one can determine directly the mixing
parameters $x_D$ and $y_D$ by studying the time evolution of the Dalitz
distribution. The signal p.d.f. for a simultaneous fit to the Dalitz and decay-time
distribution is 
\begin{equation}
\vert\langle K_S\pi^+\pi^-|D^0(t)\rangle\vert^2=
\vert\frac{1}{2}{\cal{A}}(m_-^2,m_+^2)\bigl[e^{-i\lambda_1
      t}+e^{-i\lambda_2 t}\bigr]+\frac{1}{2}\overline{\cal{A}}(m_-^2,m_+^2)\bigl[e^{-i\lambda_1 t}-
  e^{-i\lambda_2 t}\bigr]\vert^2~~,
\label{eq_10}
\end{equation}
composed of an instantaneous amplitude for $D^0$
decay, ${\cal{A}}(m_-^2,m_+^2)$, and an amplitude for the $\bar{D}^0$
decay, $\bar{\cal{A}}(m_-^2,m_+^2)$, arising due to a possibility
of mixing. They both depend on the Dalitz variables $m_-^2=m^2(K_S\pi^-)$ and
$m_+^2=m^2(K_S\pi^+)$. The dependence on the mixing parameters is hidden in 
$\lambda_{1,2}=m_{1,2}-i\Gamma_{1,2}/2$. If $CPV$ is neglected the
amplitude for $\bar{D}^0$ tagged decays is 
$\vert\langle K_S\pi^+\pi^-|\bar{D}^0(t)\rangle\vert^2(m_+^2,m_-^2,t)=
\vert\langle K_S\pi^+\pi^-|D^0(t)\rangle\vert^2(m_-^2,m_+^2,t)$. As in the
case of $B^0$ oscillation measurements the p.d.f. of Eq. (\ref{eq_10})
must be corrected to include the finite resolution on the decay time. 

Based on $\sim 500\times 10^3$ reconstructed decays with a purity of
95\% Belle obtained a good description of the Dalitz distribution
using 18 different resonant intermediate states and a non-resonant
contribution. A simultaneous fit to $m_-^2,~m_+^2$ and $t$ yielded 
mixing parameters $x_D=(0.80\pm 0.29\pm ^{0.13}_{0.16})\%$, $y_D=(0.33\pm
0.24\pm ^{0.10}_{0.14})\%$. This represents by far the most
constraining determination of $x_D$ up to date. Contour of allowed
$(x_D,y_D)$ values at 95\% C.L. is shown in Fig. \ref{fig_12} (right). 

\subsection{$CP$ violation in the system of neutral $D$ mesons}
\label{sec_32}

A general, easy to reach expectation is that possible $CPV$
in processes involving charmed hadrons must be small within the
SM. This arises due to the fact that such processes involve the first two
generations of quarks for which the elements of the CKM matrix are almost
completely real. Typical CKM factor entering both the short distance
box diagram as well as the decays to real states accessible to both, 
$D^0$ and $\bar{D}^0$, is $V_{cs}^\ast V_{us}$. Using CKM matrix
unitarity this can be expressed as $-V_{cd}^\ast V_{ud}[1+(V_{cb}^\ast
V_{ub})/(V_{cd}^\ast V_{ud})]$. Considering the small absolute value of
the second term one can see that $\arg(V_{cs}^\ast V_{us})\approx \Im((V_{cb}^\ast
V_{ub})/(V_{cd}^\ast V_{ud}))\sim 7\times 10^{-4}$. This is the
typical value of the weak phase in charmed hadron processes which
determines the size of the $CPV$ effects in the SM. For example, $CPV$
asymmetries like $A_\Gamma$ discussed below, are typically of the
order of $x_D\sin\phi$, where $\phi$ is the weak phase considered, and hence
$A_\Gamma\sim{\cal{O}}(10^{-5})$. Deviation of $|q/p|$ value from
unity, which also represents the $CP$ violation, is roughly expected to be of
the order of $\sin\phi\sim 10^{-3}$. These 
values are all below the current experimental sensitivity and any positive
experimental signature would be a clear sign of some contribution
beyond the SM. 

All three distinct types of $CP$ violation, $CPV$ in decays, in mixing
and in the interference between decays with and without mixing (see
lectures by A.J. Bevan) can in principle be present in the $D^0$
system. They are parameterized by $A_D$, $A_M$ and $\phi$ according to 
\begin{eqnarray}
\nonumber
&&\frac{A_f}{\bar{A}_{\bar{f}}}=1+\frac{A_D}{2};~~(A_D\ne 0,~CPV~{\rm
  in~decay})\\
\nonumber
&&\vert\frac{q}{p}\vert =1+\frac{A_M}{2};~~(A_M\ne 0,~CPV~{\rm
  in~mixing})\\
&&\frac{q}{p}\frac{\bar{A}_f}{A_f}=-\frac{(1+A_M/2)\sqrt{R_D}}{1+A_D/2}e^{i(\phi-\delta_f)};
~~(\phi\ne 0,~CPV~{\rm
  in~interference})~~.
\label{eq_11}
\end{eqnarray}
In the above equations $\sqrt{R_D}$ is the ratio of amplitude
magnitudes $|\bar{A}_f/A_f|$ and $\delta_f$ is the strong phase
difference between the two. 

In all mentioned mixing parameter measurements also a search for
possible $CPV$ has been performed \footnote{Note that both, mixing and
  $CPV$ searches can also be performed using the time-integrated quantities,
  for example the rate of wrong-sign semileptonic decays $D^0\to
  \bar{D}^0\to{\cal{\ell}}^-K^+\nu$ \cite{belle_klnu} or the $CP$
  asymmetry $(\Gamma(D^0\to f)-\Gamma(\bar{D}^0\to\bar{f}))/
(\Gamma(D^0\to f)+\Gamma(\bar{D}^0\to\bar{f}))$ \cite{babar_acp}.}. 

In decays to $CP$ eigenstates ($D^0\to KK,~\pi\pi$) one measures
lifetimes separately for $D^0$ and $\bar{D}^0$ tagged events. A
measurable asymmetry 
\begin{equation}
A_\Gamma=\frac{\tau(\bar{D}^0\to f_{CP})-\tau(D^0\to f_{CP})}
{\tau(\bar{D}^0\to f_{CP})+\tau(D^0\to f_{CP})}
\label{eq_12}
\end{equation}
is related to the mixing and $CPV$ parameters as \cite{bergman}
$A_\Gamma=(1/2)A_My_D\cos\phi - x_D\sin\phi$ and equals zero in the case of no
$CPV$. The measured values by Belle \cite{belle_ycp} and Babar
\cite{babar_ycp} are $A_\Gamma=(0.01\pm 0.30\pm 0.15)\%$ and
$A_\Gamma=(0.26\pm 0.36\pm 0.08)\%$, respectively. Hence there is no
sign of the $CP$ violation at the sensitivity level of around 0.3\%. 

In $D^0\to K^+\pi^-$ decays the decay time distribution is also fitted
separately for $D^0$ mesons and their anti-particles. There are six
observables, $x^{\prime\pm},~y^{\prime\pm},~R_D^\pm$, where the $\pm$
superscripts 
denote the observables for $D^0$ and $\bar{D}^0$ subsamples. They are related to
the parameters of Eq. (\ref{eq_11}) by $A_D=(R_D^+-R_D^-)/(R_D^++R_D^-)$, 
$R_M^\pm=(x^{\prime\pm 2}+y^{\prime\pm 2})/2$ and
$A_M=(R_M^+-R_M^-)/(R_M^++R_M^-)$. From such fits the results of the
search for $CPV$ in mixing and in decay are $A_M=0.1\pm
2.9,~A_D=(-2.1\pm 5.4)\%$ \cite{babar_kpi} or $A_M=0.67\pm
1.2,~A_D=(-2.3\pm 4.7)\%$ \cite{belle_kpi} (errors here include
statistical and systematic uncertainties). There is no hint of a direct
$CPV$ at the level of 5\%. 

In the $t$-dependent analysis of $D^0\to K_S\pi\pi$ Dalitz
distribution the possibility of $CPV$ is included by additional two
free parameters in the fit, $A_M$ and $\phi$. Also the direct $CPV$ can
be checked by allowing the contributions of various intermediate
states to be different for $D^0$ and $\bar{D}^0$ Dalitz
distributions. The latter was not observed within the statistical
uncertainties. Parameters of $CPV$ in mixing and interference are
found to be $A_M=-0.26\pm 0.60\pm 0.20$ and $\phi=(-0.24\pm 0.31\pm
0.09)~rad$. The contours of $(x_D,y_D)$ arising from the fit allowing for
the $CPV$ are presented in Fig. \ref{fig_12} (right). 

\subsection{Average of $D^0$ mixing parameters}
\label{sec_33}

\begin{figure}[h]
\centerline{\includegraphics[width=0.4\textwidth]{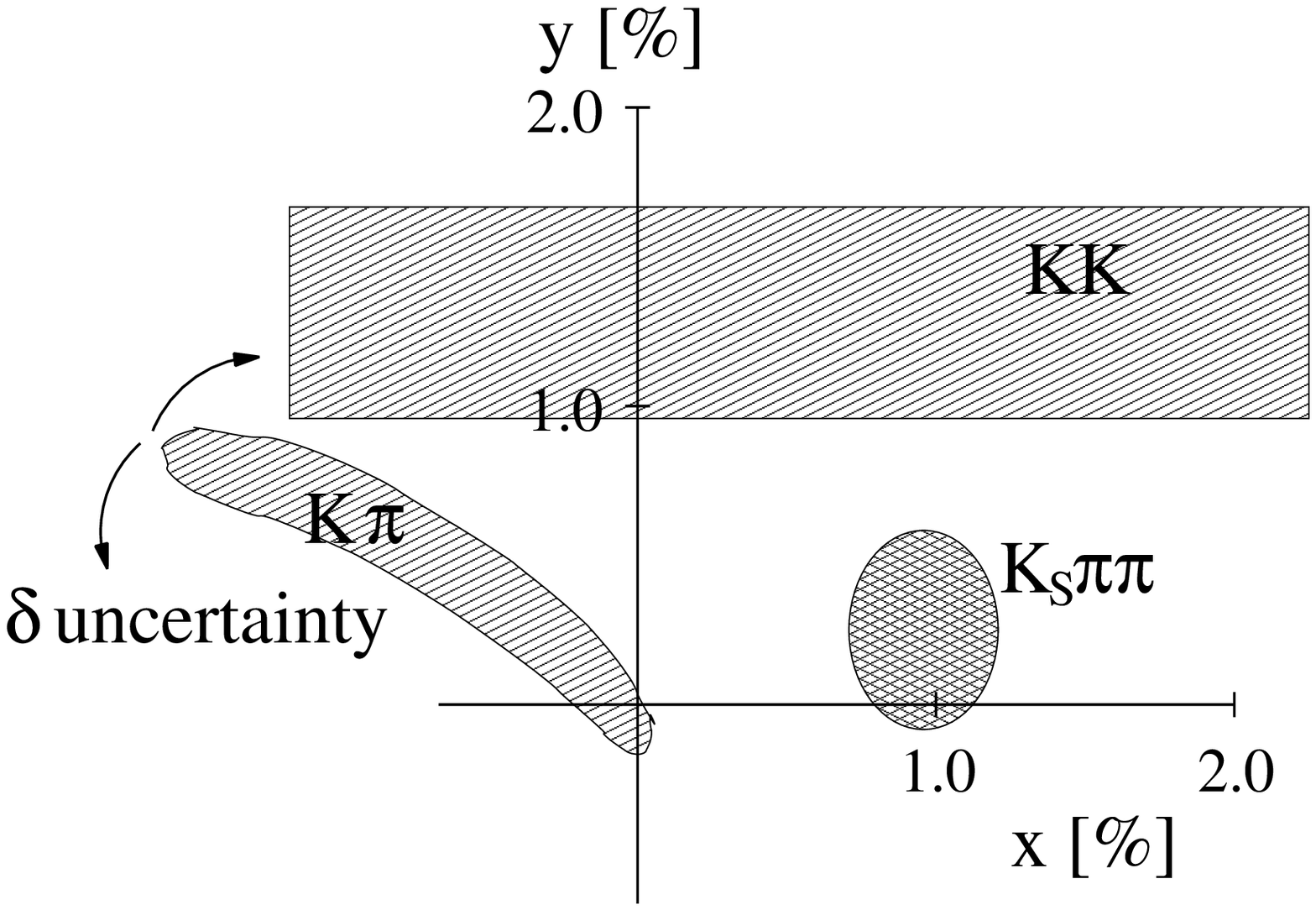}
\includegraphics[width=0.3\textwidth]{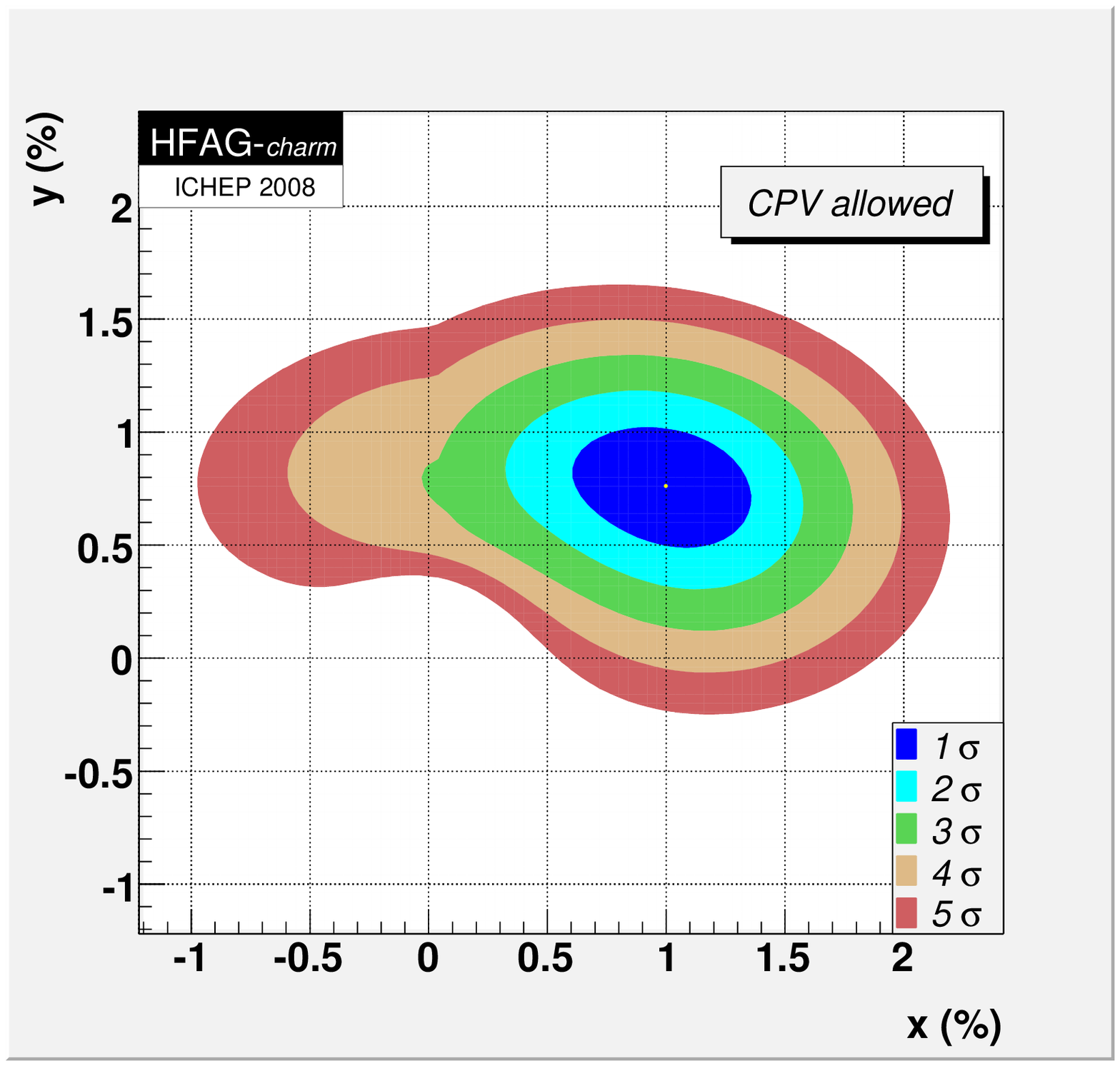}
\includegraphics[width=0.3\textwidth]{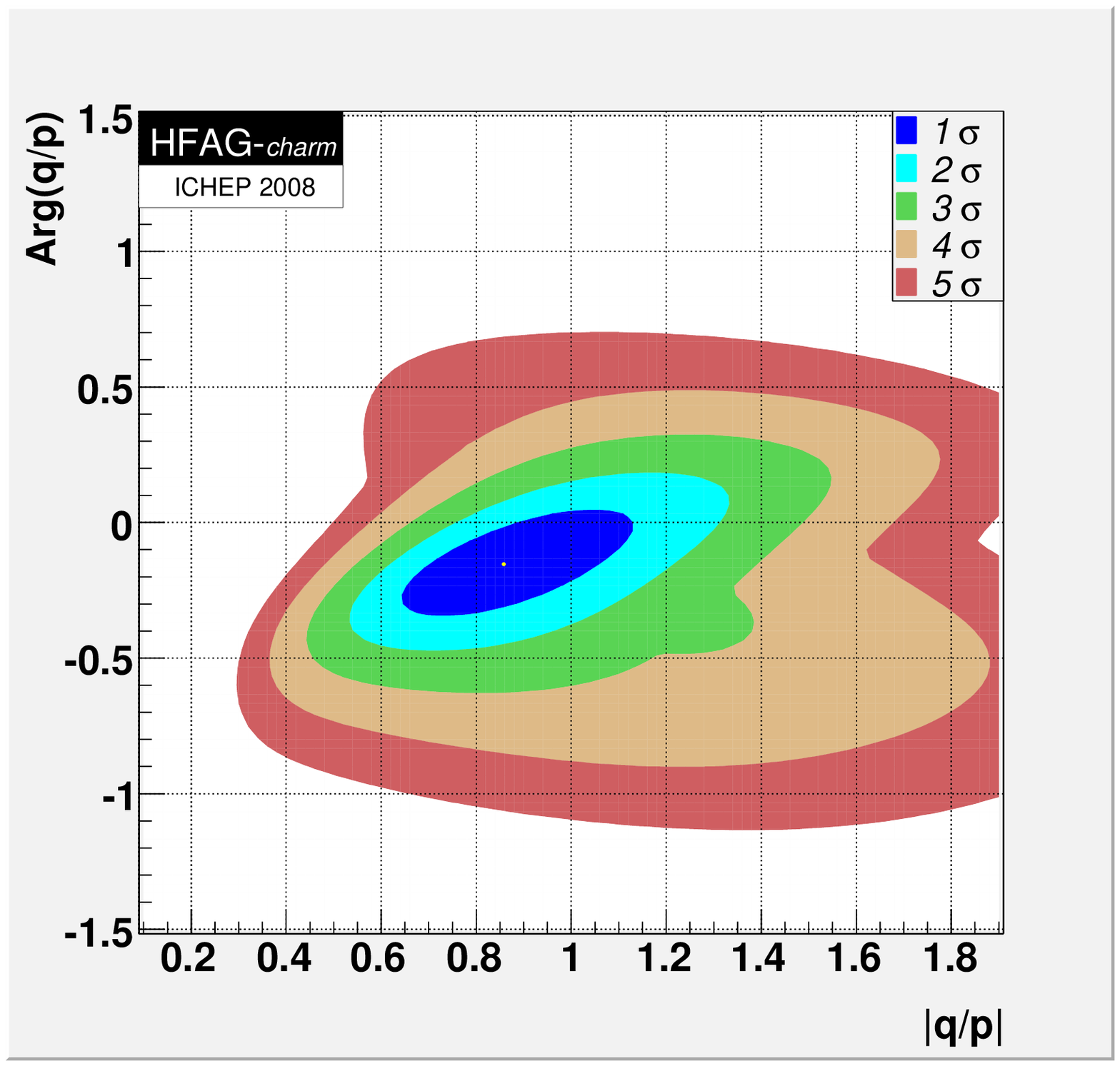}}
\caption{Left: Illustration of constraints on $(x_D,y_D)$ values arising
  from various measurements. Middle: Probability contours for $(x_D,y_D)$
  corresponding to 1-5 $\sigma$ C.L. from the average of measurements
  \cite{hfag}. Right: Probability contours for $CPV$ parameters
  $(|q/p|,\phi)$ 
  corresponding to 1-5 $\sigma$ C.L. from the average of measurements
  \cite{hfag}.}
\label{fig_13}
\end{figure}

To make conclusions arising from a variety of results on $D^0$
mixing and $CPV$ searches the Heavy Flavour Averaging Group 
performs an average of various measurements including correlations
among the measured variables \cite{hfag}. An illustration of $(x_D,y_D)$
constraints imposed by individual measurements is shown in
Fig. \ref{fig_13} (left). World average of mixing and $CPV$
parameters for the $D^0$ system is presented in Tab. \ref{tab_1}. 
\begin{table}[h]
\centerline{\begin{tabular}{|c|c|c|c|}
\hline\hline
Parameter & Value & Parameter & Value\\ \hline
$x_D$ & $(1.00\pm ^{0.24}_{0.26})\%$ & $A_D$ & $(-2.1\pm 2.4)\%$ \\
\hline
$y_D$ & $(0.76\pm ^{0.17}_{0.18})\%$ & $|q/p|$ & $0.86\pm
^{0.17}_{0.15}$ \\ \hline
$R_D$ & $(0.336\pm 0.009)\%$ & $\phi$ & $-8.8^\circ\pm
^{7.6^\circ}_{7.2^\circ}$ \\ \hline\hline
\end{tabular}}
\caption{Average of $D^0$ mixing (left) and $CPV$ (right) parameters \cite{hfag}.}
\label{tab_1}
\end{table}
The results are presented graphically in Figs. \ref{fig_13} (middle)
and \ref{fig_13} (right) as contours in $(x_D,y_D)$ and $(|q/p|,\phi)$ planes. The
mixing phenomena in the neutral $D$ meson system is firmly established,
with the mixing parameters $x_D$ and $y_D$ of the order of 1\%. The 
oscillation frequency can be compared to the values for other neutral meson
systems, $x\approx 0.8~(B^0)$,  $x_K\approx 1~(K^0)$ and  $x_s\approx
25~(B_s^0)$. Since both parameters, $x_D$ and $y_D$, appear to be
positive, it seems that the $CP$-even state of neutral charmed
mesons is shorter-lived (like in the $K^0$ system) and also heavier
(unlike in the $K^0$ system). At the moment there is no sign of $CPV$
in the $D^0$ system, at the level of one standard deviation of the world average results. 

\begin{figure}[h]
\centerline{\includegraphics[width=0.4\textwidth]{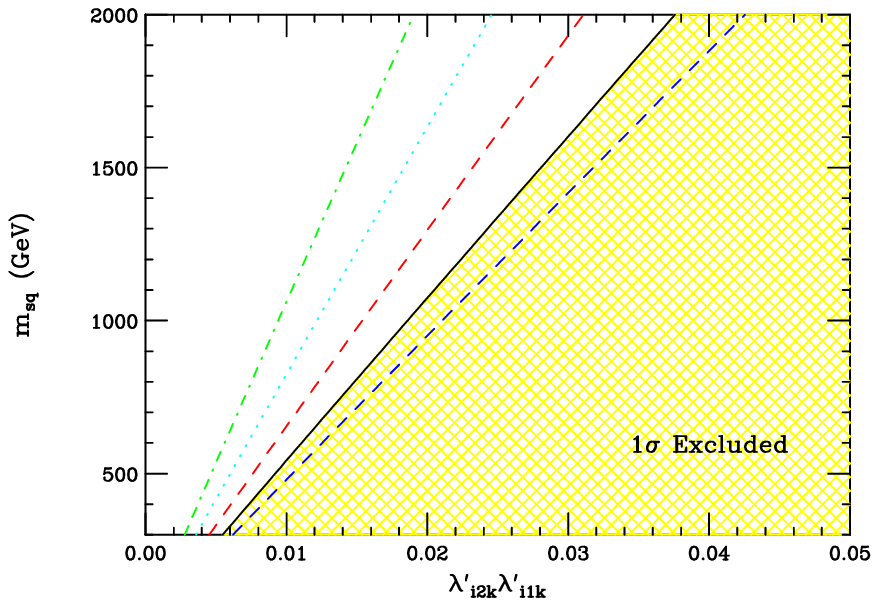}
\includegraphics[width=0.5\textwidth]{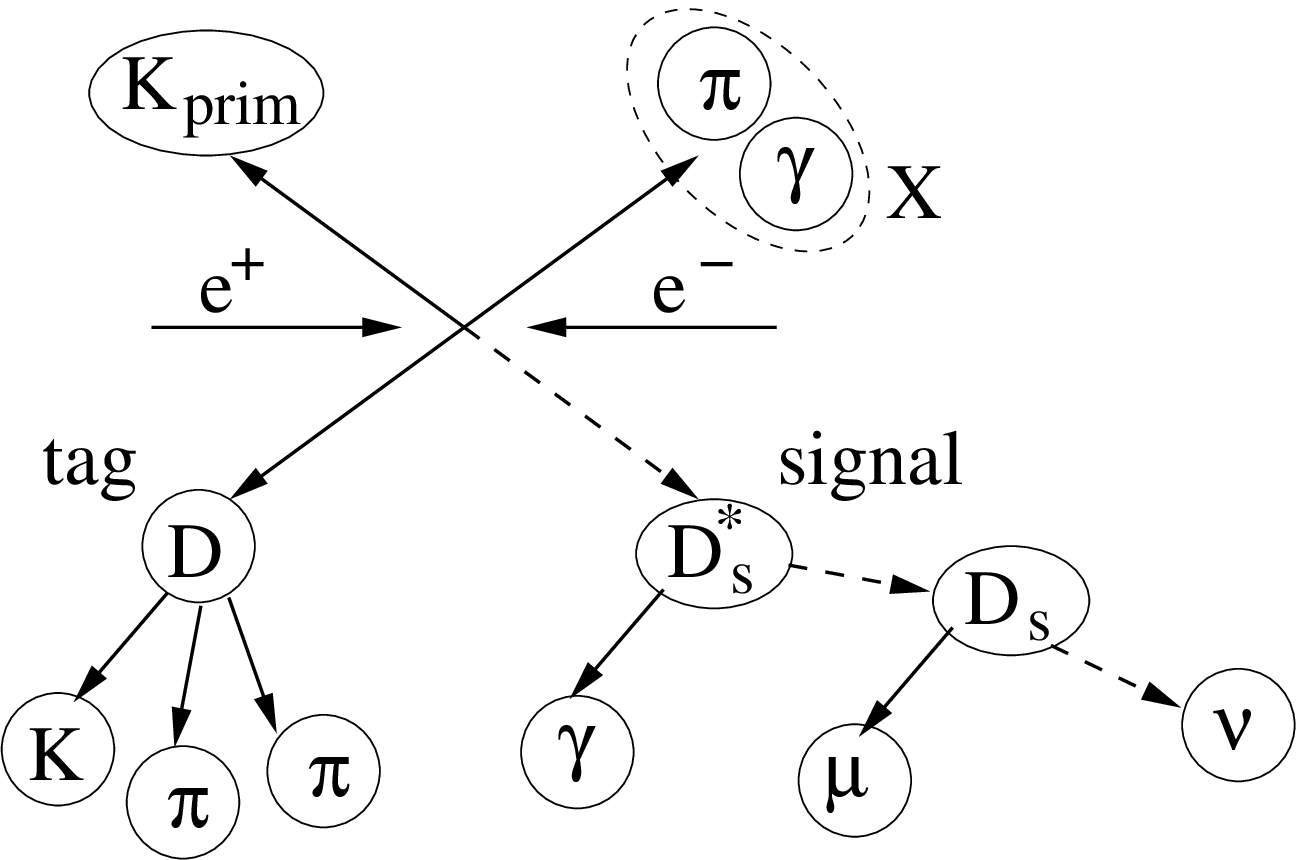}}
\caption{Left: Constraints on the values of squark mass and $R$-parity
  violating coupling constants arising from $x_D<1\%$ (hatched region
  is excluded). Dashed lines represent boundaries of the exclusion
  region 
  for $x_D<1.5\%,~0.8\%,~0.5\%$ and $0.3\%$ \cite{golowich}. Right:
  Sketch of a method to measure $Br(D_s^+\to\mu^+\nu)$
  \cite{belle_fds}. Full lines represent particles detected in the
  detector or exclusively reconstructed, and dashed lines particles reconstructed in the recoil
  (from the known momenta of incident beams and detected particles).}
\label{fig_14}
\end{figure}

Results in the $D^0$ mixing impose some stringent constraints on the
parameters of various NP models \cite{golowich}. As an example we
quote the $R$-parity violating Supersymmetry models, where
an enhancement of $x_D$ could arise from an exchange of down-like
squarks or sleptons in the loop of the box diagram. The exclusion region
of possible values of the squark mass and $R$-parity violating
coupling constants for various upper limits on $x_D$ is presented in
Fig. \ref{fig_14} (left). 
Planned Super $B$-factory, which would accumulate data corresponding
to an integrated luminosity of 50~ab$^{-1}$ (compared to the current
0.8~ab$^{-1}$ at KEKB), would of course yield
results on $D^0$ mixing and $CPV$ of much better
precision. The extrapolated accuracies are $\sigma(x_D)\sim
0.1\%,~\sigma(y_D)\sim 0.06\%,~\sigma(|q/p|)\sim 0.05$ and
$\sigma(\phi)\sim 3^\circ$. This would allow to severely constrain
relations among parameters of various NP parameters and to search for
possible $CPV$ phenomena in the region where a large number of these models
predict an observable effect. However, one should not forget the words {\it
  "Prediction is very difficult, especially of the future." (N. Bohr,
  1885 - 1962)}. 

\subsection{$D_s$ leptonic decays}
\label{sec_34}

Charmed mesons leptonic decays are analogous to the leptonic decays of
$B$ mesons (Sect. \ref{sec_22}). By measuring the rate of such
decays one would hope to determine the decay constant of the
corresponding meson,
see Eq. (\ref{eq_7}), and by that test the predictions of LQCD. Both
Belle and BaBar performed measurements of $Br(D_s^+\to\mu^+\nu)$. 
Cleo-c collaboration measured decays $D_s^+\to\tau\nu$ as well. 

In \cite{belle_fds} Belle measured the absolute branching fraction of
$D_s^+\to\mu^+\nu$ using a method illustrated in Fig. \ref{fig_14} 
(right). Events of
the type $e^+e^-\to D_s^\ast D^{\pm,0} K^{\pm,0}X$ are used, where $X$ can be any number
of additional pions from fragmentation, and up to one photon. An event is
divided into a tag side, where a full reconstruction of $D$ and
a primary $K$ meson is performed, and a signal side where the decay chain $D_s^{\ast
  +}\to D_s^+\gamma$, $D_s^+\to\mu^+\nu_\mu$ is searched for. Tag side charged and neutral $D$
mesons are reconstructed in $D\to Kn\pi$ decays. For all possible combinations of
particles in $X$, the signal side $D_s^{\ast +}$ meson is identified by
reconstruction of the recoil mass $m_{\rm rec}(DKX)$, using the known beam
momentum and four-momentum conservation. The recoil mass $m_{\rm
  rec}(Y)$ is calculated as the magnitude of the four-momentum $p_{\rm
  beams} -  p_Y$. The next step in the event reconstruction is a search for a photon for
which the recoil mass $m_{\rm rec}(DKX\gamma)$ is consistent with the
nominal mass of $D_s^+$. The sample of $D_s^+$ mesons reconstructed
using this procedure represent an inclusive sample of decays, among
which the leptonic decays are searched for. If an identified muon 
is found among the tracks so far not used
in the reconstruction, the square of the recoil mass $m^2_{\rm rec}(DKX\gamma\mu)$ is
calculated. For signal decays this mass corresponds to the
mass of the final state neutrino and hence peaks at zero. 

\begin{figure}[h]
\centerline{\includegraphics[width=0.3\textwidth]{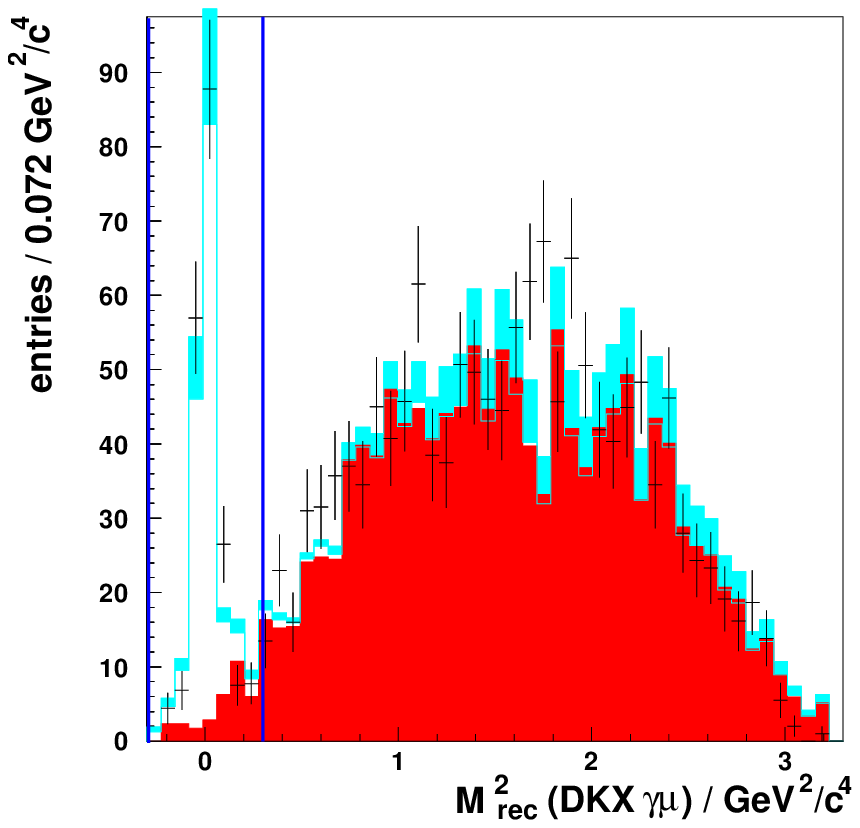}
\includegraphics[width=0.4\textwidth]{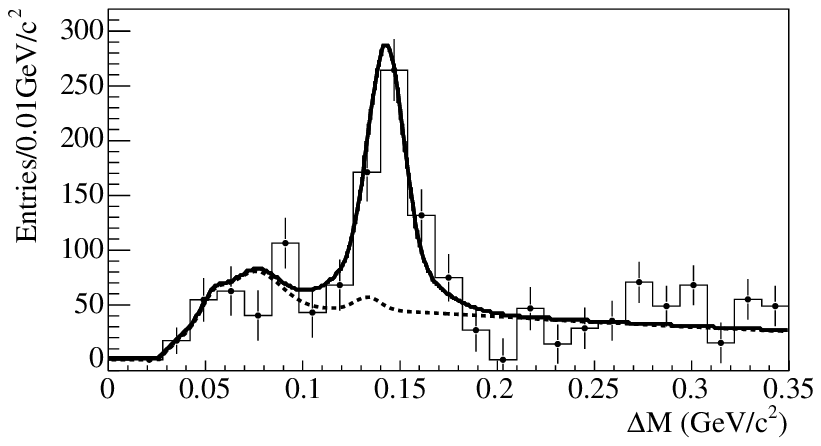}}
\caption{Left: Distribution of $m^2_{\rm rec}(DKX\gamma\mu)$ for 
  $D_s^+\to\mu^+\nu$ candidate events \cite{belle_fds}. The peak at
  zero corresponds to the signal decays. Right: $m(D_s^\ast)-m(D_s)$
  distribution for $D_s^+\to\mu^+\nu$ candidate events
  \cite{babar_fds}. Signal peaks at the nominal mass difference.}
\label{fig_15}
\end{figure}

Final
distribution of $m^2_{\rm rec}(DKX\gamma\mu)$ is shown in
Fig. \ref{fig_15} (left) where a clear signal of leptonic decays can be
  seen. Majority of background can be described using reconstructed
  $D_s^+\to e^+\nu$ decays where due to the helicity suppression no
  signal is expected. Number of reconstructed signal decays is found
  to be $N(D_s^+\to\mu^+\nu)=169\pm 16\pm 8$. Comparing to the
  number of inclusively reconstructed $D_s^+$ decays and correcting
  for the efficiency of muon reconstruction one obtains the branching
  fraction $Br(D_s^+\to\mu^+\nu)=(6.44\pm 0.76\pm 0.56)\times
  10^{-3}$. The largest contribution to the systematic uncertainty arises
  from a limited number of simulated decays used to describe the shape
  of the signal distribution. Using Eq. (\ref{eq_7}) (without the
  factor arising from the charged Higgs contribution) and the value of
  $|V_{cs}|$ as determined in a global fit to the CKM elements applying 
  the unitarity of the matrix \cite{pdg}, one determines the value of
  $D_s$ meson decay constant, $f_{D_s}=(275\pm 16\pm 12)$~MeV. 

BaBar \cite{babar_fds} used a somewhat different approach by measuring
the yield of $D_s^+\to\mu^+\nu$ relative to the $D_s^+\to\phi\pi$
decays. The branching fraction determined in this way is a relative
measurement normalized to the $Br(D_s^+\to\phi\pi)$. While this
method enables a larger statistics of the reconstructed sample it
suffers from a hard-to-estimate systematic uncertainty in the
normalization mode ($\phi\pi$ state is actually an intermediate state
of the $K^+K^-\pi^+$ final state and can be influenced by the
interference among various intermediate states). The neutrino momentum
is determined from the missing momentum in an event. The resolution is
improved by constraining the $\nu$ and the reconstructed muon momentum to
yield the nominal mass of $D_s$ meson. Fig. \ref{fig_15} (right) shows
the distribution of the reconstructed mass difference between the
$D_s^\ast$ and $D_s$ meson, where the signal of leptonic decays
consist of $489\pm 55$ events (the error is statistical
only). Calculation of the decay constant yields a value of
$f_{D_s}=(283\pm 17\pm 7\pm 14)$~MeV, where the last error is due to
the uncertainty of $Br(D_s^+\to\phi\pi)$. 

How do the measured values compare to the LQCD calculations? 
The average of absolute measurements (beside the described Belle
measurement these include measurements by Cleo-c collaboration in muon
and tau decay modes \cite{cleoc_fds}) is $f_{D_s}=(274\pm 10)$~MeV
\cite{pdg_deccons}. The recent LQCD result exhibits a huge improvement
in the accuracy compared to previous determinations of the $D_s$ meson decay constant:
$f_{D_s}=(241\pm 3)$~MeV \cite{lqcd_fds}. The discrepancy between the
two values is more than 3 standard deviations. While the fact that for
the $D^+$ decay constant experimental results confirm the calculation
(albeit within larger errors) may point to some intervention of NP
\cite{kronfeld} one should probably wait for a) confirmation of the
LQCD estimate (and especially its uncertainty) and b) more accurate
experimental measurements before making any conclusion
\footnote{For $D^+$ decays the $H^\pm$
  contribution is proportional to $(m_D^2/m_H^2)(m_d/m_c)\tan^2\beta$
  while for $D_s^+$ decays it is proportional to
  $(m_{D_s}^2/m_H^2)(m_s/m_c)\tan^2\beta$ \cite{hou_fds}. See also
  Eq. (\ref{eq_7}), where for $B^+$ decays, due to $m_b\gg m_u$, the
  correction is simply $(m_B^2/m_H^2)\tan^2\beta$.}. 

\section{Summary}

Although in the lectures we were able to present only a small
fraction of exciting physics results that arose from the $B$-factories
over almost a decade of operation, we hope the selected examples
demonstrate the following: 
\begin{itemize}
\item{$B$-factories have successfully performed precision measurements
  in identification of SM processes and determination of SM parameters, as well as a
  complement to direct NP searches that are soon to be started at the LHC;} 
\item{experimental tests in general confirm predictions of the SM,
  although several hints of discrepancies at the level of 3 standard
  deviations exist;}
\item{$B$-factories have outreached their program as foreseen at the
  startup.}
\end{itemize} 

Specifically related to the presented measurements one should note 
\begin{itemize}
\item{$B$ oscillations in conjunction with a breakthrough in $B_s$
  oscillations confirm the SM to a high accuracy;}
\item{leptonic and radiative $B$ meson decays constrain possible
  contribution of NP but large room for improvement remains for the Super
  $B$-factory;}
\item{important results in charm physics complement the results in the
  $B$ meson sector;}
\item{measurements of $D^0$ mixing and search for the $CPV$ represent
  another achieved milestone in particle physics, more precise
  measurements and theoretical predictions are needed;}
\item{$D_s$ leptonic decays may test predictions of LQCD once the
  results are confirmed.}
\end{itemize}

\subsection{Acknowledgments}
The author wishes to thank the organizers of the school for the great
summer school atmosphere and a very
pleasant stay in Dubna. Most of the measurements presented in the
lectures are the result of a splendid work of the Belle and BaBar
collaboration members and of the superb performance of the KEKB and
PEP-II accelerators. 

\vspace*{0.5cm}
\noindent The credit for the quotes used goes to:\\
Oscar Wilde, Irish playwright, poet, 1854-1900.\\
Anne Frank, Jewish girl, author of the famous diary, 1929-1945.\\
Ralph Waldo Emerson, American Poet and Essayist, 1803-1882.\\
Sir Sean Connery, Scottish actor and producer, 1930.\\
Albert Camus, Algerian-born French writer, 1913-1960.\\
%Honore de Balzac, French novelist and playwright, 1799-1850.\\
Niels Bohr, Danish physicist, 1885-1962. 

% ****************************************************************************
% BIBLIOGRAPHY AREA
% ****************************************************************************

\begin{footnotesize}
% IF YOU DO NOT USE BIBTEX, USE THE FOLLOWING SAMPLE SCHEME FOR THE REFERENCES
% ----------------------------------------------------------------------------

% ----------------------------------------------------------------------------

% IF YOU USE BIBTEX,
% - DELETE THE TEXT BETWEEN THE TWO ABOVE DASHED LINES
% - UNCOMMENT THE NEXT TWO LINES AND REPLACE 'Name_Of_Your_BibFile'

%\bibliographystyle{unsrt}
%\bibliography{Name_Of_Your_BibFile}

\begin{thebibliography}{99}
%------- replace following references ;-)
\bibitem{bevan_proc} A. Bevan, arXiv:0812.4388.
\bibitem{belle_det} A.~Abashian {\it et al.} (Belle Coll.),
  Nucl.\ Instr.\ Meth.\ A{\bf 479}, 117 (2002). 
\bibitem{kekb} S.~Kurokawa, E.~Kikutani, Nucl.\ Instr.\ Meth.\ A{\bf
  499}, 1 (2003), and other papers in this volume.
\bibitem{babar_det} B. Aubert {\it et al.} (BaBar Coll.),  Nucl.\
  Instr.\ Meth.\ A{\bf 479}, 1 (2002). 
\bibitem{cdf_det} D. Acosta {\it et al.} (CDF Coll.), Phys. Rev. D{\bf
  71}, 032001 (2005). 
\bibitem{cleoc_det} D. Peterson {\it et al.},  Nucl.\
  Instr.\ Meth.\ A{\bf 478}, 142 (2002); Y. Kubota {\it et al.} (CLEO
  Coll.), Nucl.\ Instr.\ Meth.\ A{\bf 320}, 66 (1992). 
\bibitem{pdg_cpv} D. Kirkby, Y. Nir, review {\it CP Violation in Meson
  Decays}, in C. Amsler {\it et al.}, Phys. Lett. B{\bf 667}, 1
  (2008), and references therein. 
\bibitem{belle_vtxres} H. Tajima {\it et al.}, Nucl.\ Instr.\ Meth.\ A{\bf
  533}, 370 (2004). 
\bibitem{pdg} C. Amsler {\it et al.}, Phys. Lett. B{\bf 667}, 1
  (2008). 
\bibitem{belle_dm} K. Abe {\it et al.} (Belle Coll.), Phys. Rev. D{\bf
  71}, 072003 (2005).   
\bibitem{hfag} E. Barberio {\it et al.} (HFAG), arXiv:0808.1297,
  and updates at http://www.slac.stanford.edu/xorg/hfag/
\bibitem{buras_dm} A.J. Buras {\it et al.}, Nucl. Phys. B{\bf 245},
  369 (1984). 
\bibitem{ckm_param} J. Charles {\it et al.} (CKMfitter group),
  Eur. Phys. J. C{\bf 41}, 1 (2005). 
\bibitem{ckmfit} http://ckmfitter.in2p3.fr
\bibitem{cdf_dm} A. Abulencia {\it et al.} (CDF Coll.),
  Phys. Rev. Lett. {\bf 97}, 242003 (2006). 
\bibitem{belle_btaunu_semil} I. Adachi {\it et al.} (Belle Coll.), arXiv:0809.3834.
\bibitem{babar_btaunu} B. Aubert {\it et al.} (BaBar Coll.), Phys. Rev. D{\bf 76},
  052002 (2007). 
\bibitem{belle_btaunu_had} K. Ikado {\it et al.} (Belle Coll.),
  Phys. Rev. Lett. {\bf 97}, 251802 (2006).
\bibitem{sheldon} M. Artuso, E. Barberio, S. Stone, arXiv:0902:3743.
\bibitem{pdg_deccons} See J. Rosner, S. Stone, review {\it Decay
  Constants of Charged Pseudoscalar Mesons}, in C. Amsler {\it et
  al.}, Phys. Lett. B{\bf 667}, 1
  (2008), and references therein.
\bibitem{lqcd_fb} A. Gray {\it et al.} (HPQCD Coll.),
  Phys. Rev. Lett. {\bf 95}, 212001 (2005).  
\bibitem{belle_bsg} K. Abe {\it et al.} (Belle Coll.),
  arXiv:0804.1580. 
\bibitem{babar_bsg} B. Aubert {\it et al.} (BaBar Coll.),
  Phys. Rev. D{\bf 72}, 052004 (2005). 
\bibitem{misiak_bsg} M. Misiak {\it et al.}, Phys. Rev. Lett. {\bf
  98}, 022002 (2007).  
\bibitem{hurth_bsll} U. Egede {\it et al.}, arXiv:0807.2589, and
  references therein. 
\bibitem{ali_bsll} A. Ali {\it et al.}, Phys. rev. D{\bf 66}, 034002
  (2002). 
\bibitem{babar_bsll} B. Aubert {\it et al.} (BaBar Coll.),
  arXiv:0804.4412. 
\bibitem{belle_bsll} I. Adachi {\it et al.} (Belle Coll.), arXiv:0810.0335.
\bibitem{kamenik} T. Hurth {\it et al.}, arXiv:0807.5039.
\bibitem{burdman} G. Burdman, I. Shipsey, Ann. Rev. Nucl. Sci. {\bf
  53}, 431 (2005). 
\bibitem{dmix_th} I.I. Bigi, N. Uraltsev, Nucl. Phys. B{\bf 592}, 92
  (2001); A.F. Falk {\it et al.}, Phys. Rev. D{\bf 69}, 114021
  (2004). 
\bibitem{bergman} S. Bergmann {\it et al.}, Phys. Lett. B {\bf 486},
  418 (2000). 
\bibitem{belle_ycp}   M. Stari\v c {\it et al.} (Belle Coll.), Phys.\ Rev.\
  Lett.\ {\bf 98}, 211803 (2007). 
\bibitem{babar_kpi}  B. Aubert {\it et al.} (BaBar Coll.), Phys.\ Rev.\
  Lett.\ {\bf 98}, 211802 (2007). 
\bibitem{babar_ycp} B. Aubert {\it et al.} (BaBar Coll.), Phys.\
  Rev. D{\bf 78}, 011105 (2008). 
\bibitem{belle_kspipi} L.M. Zhang {\it et al.} (Belle Coll.), Phys.\
  Rev.\ Lett.\ {\bf 99}, 131803 (2007).
\bibitem{belle_klnu} U. Bitenc {\it et al.} (Belle Coll.),
  Phys. Rev. D{\bf 77}, 112003 (2008). 
\bibitem{babar_acp}  B. Aubert {\it et al.} (BaBar Coll.), Phys.\ Rev.\
  Lett.\ {\bf 100}, 061803 (2008).
\bibitem{belle_kpi}  L.M. Zhang {\it et al.} (Belle Coll.), Phys.\ Rev.\
  Lett.\ {\bf 96}, 151801 (2006). 
\bibitem{golowich} E. Golowich {\it et al.}, Phys.\ Rev. D{\bf 76},
  095009 (2007).
\bibitem{belle_fds} L. Widhalm {\it et al.} (Belle Coll.),
  Phys. Rev. Lett. {\bf 100}, 241801 (2008). 
\bibitem{babar_fds} B. Aubert {\it et al.} (BaBar Coll.),
  Phys. Rev. Lett. {\bf 98}, 141801 (2007). 
\bibitem{lqcd_fds} E. Follana {\it et al.} (HPQCD and UKQCD Coll.),
  Phys. Rev. Lett. {\bf 100}, 062002 (2008). 
\bibitem{cleoc_fds} M. Artuso {\it et al.} (Cleo-c Coll.),
  Phys. Rev. Lett. {\bf 99}, 071802 (2007); K.M. Ecklund {\it et al.} (Cleo-c Coll.),
  Phys. Rev. Lett. {\bf 100}, 161801 (2008). 
\bibitem{kronfeld} B.A. Dobrescu, A.S. Kronfeld, Phys. Rev. Lett. {\bf
  100}, 241802 (2008). 
\bibitem{hou_fds} W.S. Hou, Phys. Rev. D{\bf 48}, 2342 (1993). 

\end{thebibliography}
% example of Name_Of_Your_BibFile.bib
% @Article{Turcato:2006ch,
%      author    = "Turcato, M.",
%  collaboration = "ZEUS and H1",
%      title     = "Lepton flavour violation and charmonium physics at HERA",
%      journal   = "Nucl. Phys. Proc. Suppl.",
%      volume    = "162",
%      year      = "2006", 
%      pages     = "283-287",
%      SLACcitation  = "%%CITATION = NUPHZ,162,283;%%"
% }
% 
% @Unpublished{Gogitidze:2007du,
%      author    = "Gogitidze, N.",
%  collaboration = "H1", 
%      title     = "Prompt photons and particle momentum distributions at
%                   HERA", 
%      year      = "2007",
%      note    = "hep-ex/0701033",
%      SLACcitation  = "%%CITATION = HEP-EX 0701033;%%"
% }

\end{footnotesize}

% ****************************************************************************
% END OF BIBLIOGRAPHY AREA
% ****************************************************************************

\end{document}